\documentclass[twocolumn,superscriptaddress,showpacs,prd,aps,amsmath,amssymb,nofootinbib]{revtex4-1}
\usepackage{natbib}

\usepackage{graphicx,color}
\usepackage{amsmath,amssymb}
\usepackage{verbatim}
\usepackage{hyperref}
\usepackage{float}
\usepackage{wasysym}
\usepackage{amssymb,graphicx}
\usepackage{epsfig}
\usepackage{psfrag}
\usepackage{dsfont}
\usepackage{amsfonts}
\usepackage{mathrsfs}
\usepackage{multirow}
\usepackage{times}

\newcommand{\bra}[1]{\left( #1 \right)}

\newcommand{\mnras}{Mon.\ Not.\ R.\ Astron.\ Soc.\ }

\begin{document}

\title{Disks Around Merging Binary Black Holes: From GW150914 to Supermassive Black Holes}
\author{Abid Khan}
\affiliation{Department of Physics, University of Illinois at
  Urbana-Champaign, Urbana, IL 61801}
\author{Vasileios Paschalidis}
\affiliation{Theoretical Astrophysics Program, Departments of Astronomy and Physics, University of Arizona,
 Tucson, AZ 85721}
\affiliation{Department of Physics, Princeton University,
  Princeton, NJ 08544}
\author{Milton Ruiz}
\affiliation{Department of Physics, University of Illinois at
  Urbana-Champaign, Urbana, IL 61801}
\author{Stuart L. Shapiro}
\affiliation{Department of Physics, University of Illinois at
  Urbana-Champaign, Urbana, IL 61801}
\affiliation{Department of Astronomy \& NCSA, University of
  Illinois at Urbana-Champaign, Urbana, IL 61801}


\begin{abstract}
  We perform magnetohydrodynamic simulations in full general
  relativity of disk accretion onto nonspinning black hole binaries
  with mass ratio $q=29/36$. We survey different disk models which differ
  in their scale height, total size and magnetic field to quantify the
  robustness of previous simulations on the initial disk
  model. Scaling our simulations to LIGO GW150914 we find that such
  systems could explain possible gravitational wave and
  electromagnetic counterparts such as the Fermi GBM hard X-ray signal
  reported 0.4s after GW150915 ended. Scaling our simulations to
  supermassive binary black holes, we find that observable flow
  properties such as accretion rate periodicities, the emergence of
  jets throughout inspiral, merger and post-merger, disk temperatures,
  thermal frequencies, and the time-delay between merger and the boost
  in jet outflows that we reported in earlier studies display only
  modest dependence on the initial disk model we consider here.
\end{abstract}

\maketitle

\section{Introduction}

Accreting black holes are central in explaining a range of high-energy
astrophysical phenomena that we observe in our Universe, such as X-ray
binaries, active galactic nuclei (AGN), and quasars. Recently,
substantial theoretical and observational effort has gone into
understanding accretion onto binary black holes and the emergent
electromagnetic (EM) signatures these systems may generate, because
they are anticipated to exist at the centers of distant AGNs and
quasars (see, e.g.,~\cite{gold} for a summary of relevant work prior
to 2014, and~\cite{Zilh2015PhRvD..91b4034Z,Farris2015MNRAS.447L..80F,
  Nguyen2016ApJ...828...68N,Kulkarni2016MNRAS.456.3964K,
  Rafikov2016ApJ...827..111R,Bowen2017ApJ...838...42B,
  Ryan2017ApJ...835..199R,Haiman2017PhRvD..96b3004H,
  DOrazio2017arXiv170702335D,Kharb2017arXiv170906258K,Bowen:2017oot}
and references therein for some more recent work). The bulk of the
research so far has focused on supermassive black holes binaries and
about 150 candidate accreting supermassive black hole binaries have
been identified in quasar surveys~\cite{2015MNRAS.453.1562G,
  Charisi2016MNRAS.463.2145C,graham}. Depending on the physical
properties of the above systems, such as the mass ratio and orbital
period, some of these candidates may be in the gravitational-wave
driven regime~\cite{DOrazio:2015tay}.  However, in addition to
accreting supermassive binary black holes, there may exist black hole
binaries of a few tens of solar masses that could be accreting matter
from a circumbinary disk. This scenario has attracted a lot of
attention recently because of the direct detection of gravitational
waves (GWs).

In particular, on September 14, 2015, the LIGO and Virgo
collaborations made the first direct detection of a GW signal -- event
GW150914~\cite{abbot}. GW150914 was entirely consistent with an
inspiraling and merging binary black hole (BHBH) in vacuum as
predicted by the theory of general relativity (GR). This detection
provided the best evidence yet for the existence of black
holes. Meanwhile, the detection of a transient EM signal (event
GW150914-GBM) at photon energies $>50$ keV that lasted 1 s and
appeared 0.4 s after the GW signal was reported in~\cite{connaughton},
who used data from the Gamma-ray Burst Monitor (GBM) aboard the Fermi
satellite. The Fermi GBM satellite was covering 75\% of the
probability map associated with the LIGO localization event in the
sky, thus this signal could be a chance coincidence. 
We note that while the detection of GW150914-GBM has been
controversial (see e.g.~\cite{Xiong:2016ssy,Greiner:2016dsk}), a
recent reanalysis of the data reported in~\cite{connaughton} finds no
reasons to question it \cite{connaughton2018}. On the
other hand, an MeV-scale EM signal lasting for 32 ms and occurring
0.46 s before the third GW detection made by the LIGO detectors (event
GW170104) -- also consistent with a BHBH~\cite{2017PhRvL.118v1101A} --
was reported in~\cite{Verrecchia2017arXiv170600029V}, who used data
from the AGILE mission. While these candidate counterpart EM signals
were not confirmed by other satellites operating at the same time,
they still excited the interest of the community and several ideas
about how to generate accretion disks onto ``heavy'' black holes, such
as those that LIGO detected, have been proposed.

Circumbinary disks were proposed because inspiraling and merging BHBHs
in vacuum do not generate EM signals, but they may do so in the
presence of ambient gas, such as binaries residing in
AGNs~\cite{bartos,stone} or those with gas remaining from their
stellar progenitors~\cite{perna,loeb,murase,li}. There are multiple
papers that suggest possible connections between LIGO GW150914 and
GW150914-GBM, see, e.g.,~\cite{loeb, janiuk, bejger}.  Sources that
could generate both the GWs and the hard X-rays reported include
binary black hole-neutron stars, binary black hole-massive stars, and
rapidly rotating massive stars. In~\cite{fedrow, dal, ackermann}
constraints on the resulting BHBH and ambient gaseous environment are
discussed. However, some of these constraints may not account for all
the physical processes involved, thus accretion onto BHBHs under a
wide number of scenarios may still be relevant not only for
supermassive black holes found in AGNs, but also for LIGO's heavy
black holes. Future ``multimessenger'' observations of such systems
will better inform us regarding the existence of accreting heavy
BHBHs.

In previous work~\cite{farris2,gold,gold2} we initiated
magnetohydrodynamic (MHD) studies of nonspinning binary black holes
accreting from a circumbinary disk in full GR. We modeled both the
binary-disk predecoupling and post-decoupling phases. We investigated
the effects of magnetic fields and the binary mass ratio, and
discovered that these systems launch magnetically driven jets (even
though the black holes were nonspinning) whose Poynting luminosity is
of order 0.1\% of the accretion power. Therefore, these systems could
serve both as radio, X-ray and gamma-ray engines for supermassive
BHBHs in AGNs and as gamma-ray-burst-like engines for LIGO-observed
BHBH masses, if in the latter case the hyperaccreting phase associated
with the merger lasts for $O(1\rm s)$.  We also discovered that $\sim
2-5(M/10^8M_\odot)(1+z)$ d [or $\sim 0.1-0.3(M/65M_\odot)(1+z)$ s]
following merger, there is a significant boost both in the Poynting
luminosity and the efficiency for converting accretion power to EM
luminosity. The time delay in units of M between merger and the jet
luminosity boost, as well as the magnitude of the boost, depends on
the binary mass ratio. In particular, more equal-mass binaries exhibit
a longer time delay and stronger boost. The origin of this boost is
currently unclear. It could be due to the fact that, prior to merger,
the black holes in our simulations were nonspinning, while after
merger a single, spinning black hole formed. It could also be due to
an increase of magnetic flux accreted onto the BHs prior to and after
merger, as the binary tidal torques modify the accretion flow -- the
higher the mass ratio the weaker the tidal torques and the more
magnetized matter is present in the vicinity of the binary prior to
merger. It has never been explored whether this time delay depends on
the initial disk model. For example, one could anticipate some
dependence on the disk thickness because the binary tidal torques
depend on the disk thickness, as well (see e.g.~\cite{kocsis, kocsis2, 
  Armitage1538-4357-567-1-L9,Liu2010PhRvD..82l3011L,2012MNRAS.427.2660K,
  Kocsis:2012ui,Shapiro:2013qsa}).

In this paper we consider a black hole binary with mass ratio
$q=29/36$ --
motivated by the inferred value of the mass ratio of GW150914 -- and
consider different initial disk models to explore their impact on the
emergence of jets, the outgoing Poynting luminosity, the presence of
periodicities in the accretion rate, and the time delay between merger
and the boost in the jet luminosity that we discovered before. As in
our previous studies we ignore the disk self-gravity and detailed
microphysics, which then allows us to scale the binary black hole mass
and disk densities to arbitrary values. Thus, our results can be
applied to both supermassive black hole binaries and LIGO/Virgo black
hole binaries.

Scaling our simulations to the GW150914 mass scale, we find that
magnetized accretion onto binary black holes could explain both the
GWs detected from this system and the EM counterpart GW150915-GBM
reported by the Fermi GBM team 0.4s after LIGO's GW150915
detection. We also find that flow properties such as the emergence of
jets, accretion rate periodicities, temperatures, thermal frequencies,
and the time-delay between merger and the boost in jet outflows that
we reported in our earlier studies display only modest dependence on
the disk model.

The rest of the paper is structured as follows: In
Sec. \ref{sec:overview} we provide a qualitative overview of the
evolution of a BHBH -- accretion disk system to motivate some of the
parameters we choose in our simulations. In Sec.~\ref{sec:methods} we
summarize the methods we adopt for generating initial data and for the
dynamical evolution. A general description of our simulation results
are reported in Sec.~\ref{sec:results}. In Sec.~\ref{sec:astro} we
discuss the astrophysical implications of our findings, both in the
context of possible EM counterparts to LIGO/Virgo GW sources and in
the context of luminous AGNs. We summarize our conclusions in
Sec.~\ref{sec:conclusions}. Geometrized units, where $G=1=c$, are
adopted throughout unless stated otherwise.


\section{Qualitative Overview}
\label{sec:overview}        

The evolution of a circumbinary disk is roughly composed of three
phases: (i) The early inspiral \emph{predecoupling} phase, during
which the disk viscous timescale ($t_{\rm vis}$) is shorter than the
GW timescale ($t_{\rm GW}$), and the quasistationary disk tracks the
BHBH as it slowly inspirals; (ii) the \emph{postdecoupling} phase,
during which $t_{\rm vis} > t_{\rm GW}$ and the binary decouples from
the disk and runs away, leaving the disk behind with a subsequent
decrease in the accretion rate; (iii) The post-merger or
\emph{re-brightening/afterglow} phase during which the disk begins to
refill the partial hollow left ``behind'' the BHBH and accretion ramps
up onto the spinning remnant BH.

The disk structure at decoupling determines the subsequent evolution
and the EM emission. A rough estimate of the decoupling radius for our
simulations is obtained by equating the viscous timescale $t_{\rm
  vis}$ to the GW inspiral timescale $t_{\rm GW}$. The former is
approximately given by
\[t_{\rm vis} \sim \frac{2}{3}\frac{R_{in}^{2}}{\nu},\]
where, $\nu$ is the effective kinematic viscosity induced by MHD
turbulence, and $R_{in}$ is the radius of the disk inner edge. We can
fit the viscosity to an $\alpha$-disk law for an analytic estimate
(see~\cite{slshapir}):
\[\nu(R) = \frac{2}{3}\alpha\left(H/R\right)^{2}\left(MR\right)^{1/2},\]
where $H$ is the thickness of the disk. The GW inspiral timescale is
given roughly by
\[t_{GW} \approx \frac{5}{64}\frac{a^{4}}{M^{2}\mu},\]
where $a$ is the binary separation, $M$ is the total BHBH
mass, and $\mu$ is the reduced mass.  Denoting the inner radius of the
disk as $R_{in} = \beta a$, we find the decoupling radius $a_{d}$ to
be
\[\frac{a_{d}}{M} \approx 13
		\left(\frac{\beta}{1.4}\right)^{3/5}\left(\frac{M}{4\mu}\right)^{-2/5}
                \left(\frac{\alpha}{0.13}\right)^{-2/5}\left(\frac{H/R}{0.2}\right)^{-4/5}, \]
where the normalizations are based on typical values the parameters
obtain in our canonical simulation described below. This estimate
suggests that when choosing initial conditions the binary orbital
separation should be larger than $13M$. In our simulations we choose
an initial separation of $\sim 14M$ and find that decoupling occurs at
orbital separations $\sim 10M$, consistent with the above
estimate. This allows the BHBH to undergo $\sim 10$ nearly-circular
orbits before reaching the decoupling radius $a_{d}$, which in turn,
allows the inner disk to relax before decoupling, yielding a
quasistationary accretion flow.

Although the above picture regarding the evolution of the matter in
the circumbinary disk allows us to estimate the decoupling radius, we
point out that recent numerical simulations have challenged it (see
e.g. \cite{gold,gold2,Farris2015MNRAS.447L..80F,Duffell:2014jma,Edgar:2007qr}).
In particular, it has been found that dense spiral accretion streams
are attached onto the BHs during the entire inspiral. Nevertheless, in
fully general relativistic simulations of thick accretion
disks~\cite{Farris2015MNRAS.447L..80F,gold2}, the accelerated inspiral
during the late stages causes the accretion rate to plummet by factors
of several. This phase can be taken to signal the "decoupling" of the
binary from the disk and matches the prediction of the last equation.


\section{Methods}\label{sec:methods}

\subsection{Metric Initial Data}

We prepare the spacetime metric initial data using the {\tt
  TwoPunctures} code~\cite{Ansorg2004PhRvD..70f4011A} and choose the
puncture momenta such that the BHBH is initially on a quasi-circular
orbit at coordinate orbital separation of $\sim 14M$. We choose the
puncture bare masses such that the ratio of irreducible masses is
$q=29/36$, and set the spins to zero ($a_{1}/M_{1} = 0 =
a_{2}/M_{2}$), which are both consistent with the mass ratio and
effective spin $\chi = M_{1}(a_{1}/M_{1}) +
M_{2}(a_{2}/M_{2})\approx0$ reported for GW150914. The initial orbital
period of the binary is $\sim 368M$, and the total number of orbits
until BHBH merger is about 16, (see Tab.~\ref{tab:t1} for a summary of
the BHBH properties).

\subsection{Matter and Magnetic-field Initial Data}

We set up a magnetized accretion disk as in~\cite{gold}. The initial
disk configurations we consider would be equilibrium solutions around
{\it single} nonspinning black holes which have the same mass as the
ADM mass of the BHBH, if the disk was not magnetized.  In particular,
we use the same family of hydrodynamic equilibrium disk models around
single black holes discussed in~\cite{farris3, villiers}, which we
construct by adopting a $\Gamma$-law equation of state (EOS),
$P=(\Gamma-1)\,\epsilon\,\rho_0$.  The $\Gamma$-law EOS is also
adopted for the dynamical evolution and allows for shock heating. Here
$\epsilon$ is the specific internal energy and $\rho_0$ the rest-mass
density. In all our models we set $\Gamma=4/3$, which is appropriate
when thermal radiation pressure dominates and allows for comparisons
with our earlier simulations reported in~\cite{gold}. More detailed
microphysics must be incorporated to replace this simple EOS with a
more realistic model, but the adopted $\Gamma$-law EOS suffices for
our initial survey in which we wish to understand the basic MHD flow
and gross EM luminosity features.

\begin{figure}
\centering
\includegraphics[scale=0.32]{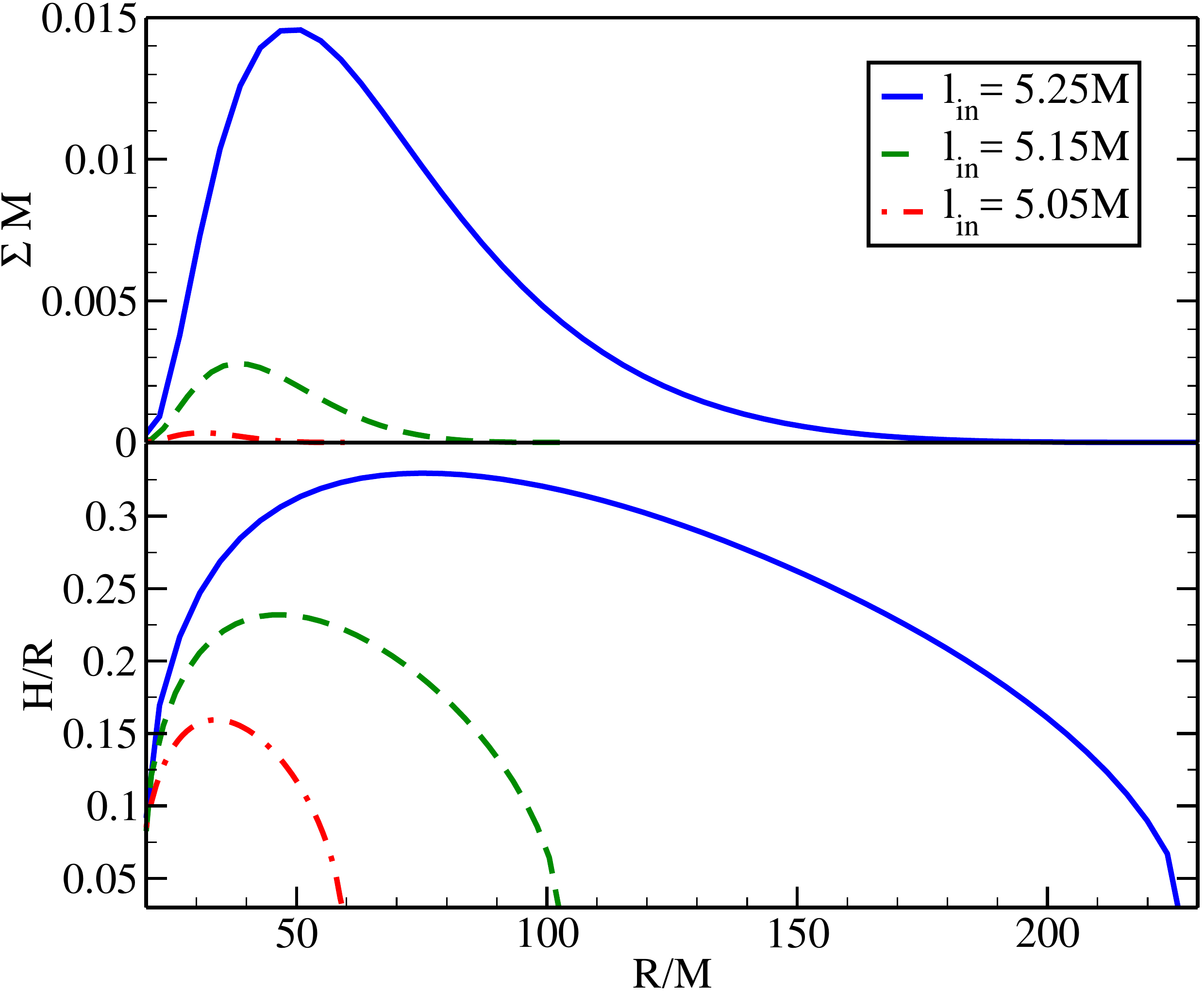}
\caption{Plots of surface density profile $\Sigma(R)$ and disk scale
  height $H/R$ vs. cylindrical coordinate radius $R$ for the three
  disk models. These initial disk models belong to a one-parameter
  family of models in which the parameter that varies is the specific
  angular momentum parameter ($l_{\rm in}$) at the disk inner edge,
  whose initial radius we fix at $20M$. For cases A, B, and C we have
  $l_{\rm in}=5.25M$ (solid lines), $l_{\rm in}=5.15M$ (dashed lines),
  and $l_{\rm in} = 5.05M$ (dot-dashed lines), respectively (see
  Tab.~\ref{tab:t2}).\label{fig:cases}}
\end{figure}

\begin{center}
  \begin{table}[th]
    \caption{Two puncture initial data parameters for the BHBH
      spacetime. Columns show the mass ratio ($q$), defined as the ratio of
      the BH irreducible masses ($M_{\rm irr}^{i}$, $i = 1,2$), the
      binary coordinate separation ($a/M$), the binary angular
      velocity $\Omega M$, and total ADM angular momentum $J/M^{2}$,
      where $M$ is the total ADM mass. Both BHs are
      nonspinning.  \label{tab:t1}}
    \begin{tabular}{cccc} 
      \hline
      \multicolumn{4}{c}{BHBH Initial Data Parameters}\\
      \hline
      \hline
      $q=M_{\text{irr}}^{2}/M_{\text{irr}}^{1}$ & $a/M$ & $\Omega M$ & $J/M^2$\\
      \hline
      29/36 & 13.8 & $1.71\times 10^{-2}$ & 1.07\\
      \hline
    \end{tabular}
  \end{table}
\end{center}

We consider three cases for the disk model by varying the geometric
thickness of the disk. Figure~\ref{fig:cases} shows the difference
between the different disk models via a plot of the initial surface
density $\Sigma(R) = \int_{z\geq0}\rho_{0}u^{t}\sqrt{-g}\,dz$, where
$R$ is the cylindrical coordinate radius, and the scale height
$H(R)=\Sigma(R)/\rho_{0}(z=0)$. In Tab.~\ref{tab:t2} we list basic
properties of these initial disk models.

By neglecting the disk self-gravity and adopting a $\Gamma$-law EOS,
we have the freedom to scale the binary ADM mass and disk densities to
any value. But the scaling remains physically valid only when the disk
mass $M_{\rm disk}$ is much smaller than the BHBH ADM mass $M$. This
scale freedom makes our calculations relevant both for AGN
environments and for environments that can explain the observed Fermi
GBM luminosity of $\sim 10^{49}$ erg/s reported in~\cite{connaughton}
(see Appendix~\ref{sec:appendixA} for a discussion of how this scaling
works for several physical quantities).

\begin{center}
  \begin{table}[h] 
    \caption{Here $l_{\rm in} = -u_{\phi}/u_{t}$ is the specific
      angular momentum parameter at the inner radius of the disk,
      $(H/R)_{\text{max}}$ is the maximum disk scale height, and
      $R_{\rm out}/M$ is the outer radius of the disk. In all cases,
      the inner radius of the disk is $R_{\rm in}/M =
      20$.  \label{tab:t2}}
      \begin{tabular}{cccc} 
        \hline
        \multicolumn{4}{c}{Initial Disk Parameters}\\
        \hline
        \hline
        Case & $l_{\rm in}/M$ & $(H/R)_{\rm max}$ & $R_{\rm out}/M$\\
        \hline
        A & 5.25 & 0.33  & 250\\
        B & 5.15 & 0.22  & 100\\
        C & 5.05 & 0.14  & 60 \\
        \hline
      \end{tabular}
\end{table}
\end{center}

\begin{figure*}
\centering
\includegraphics[scale=0.29]{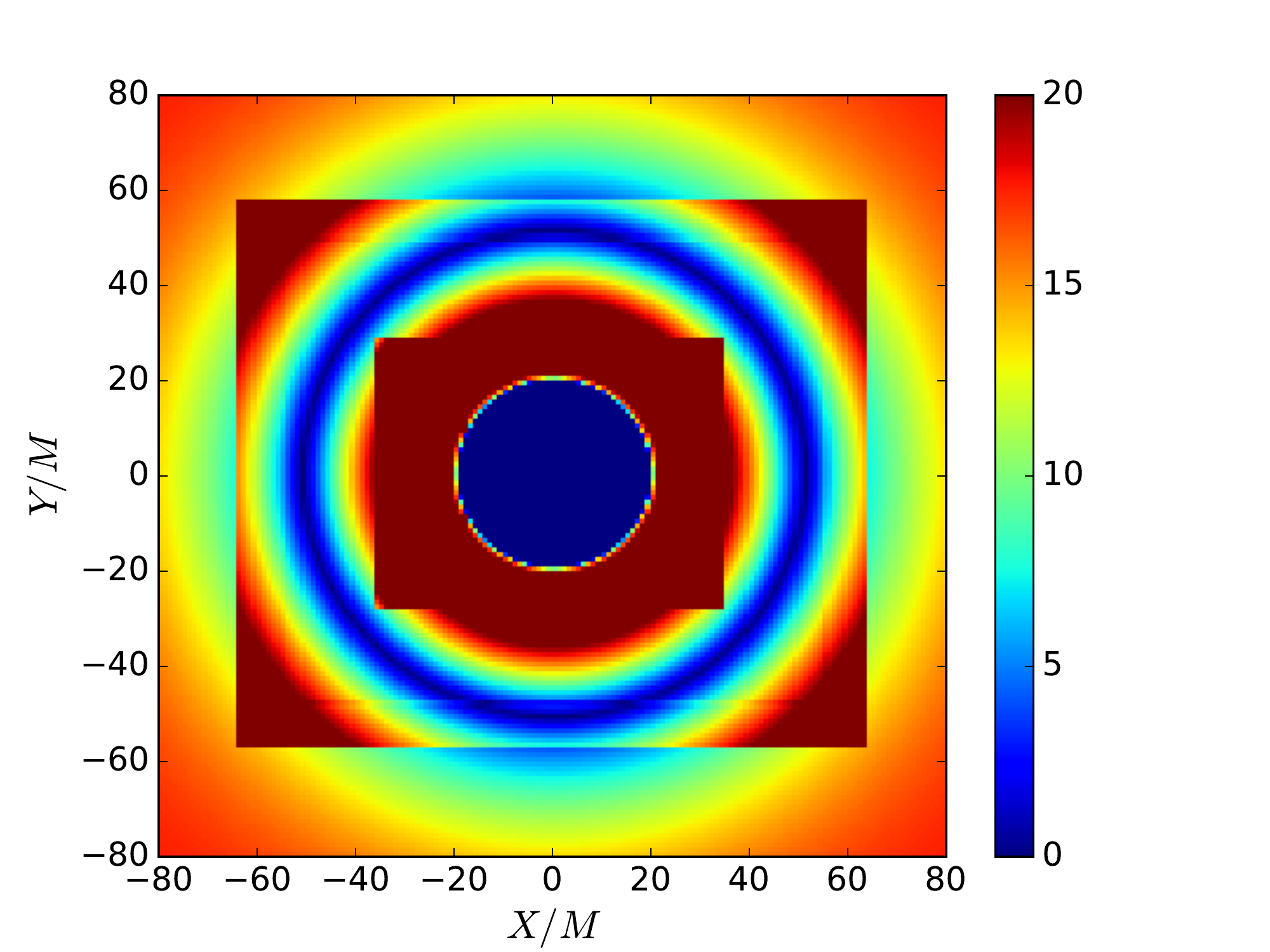}
\includegraphics[scale=0.29]{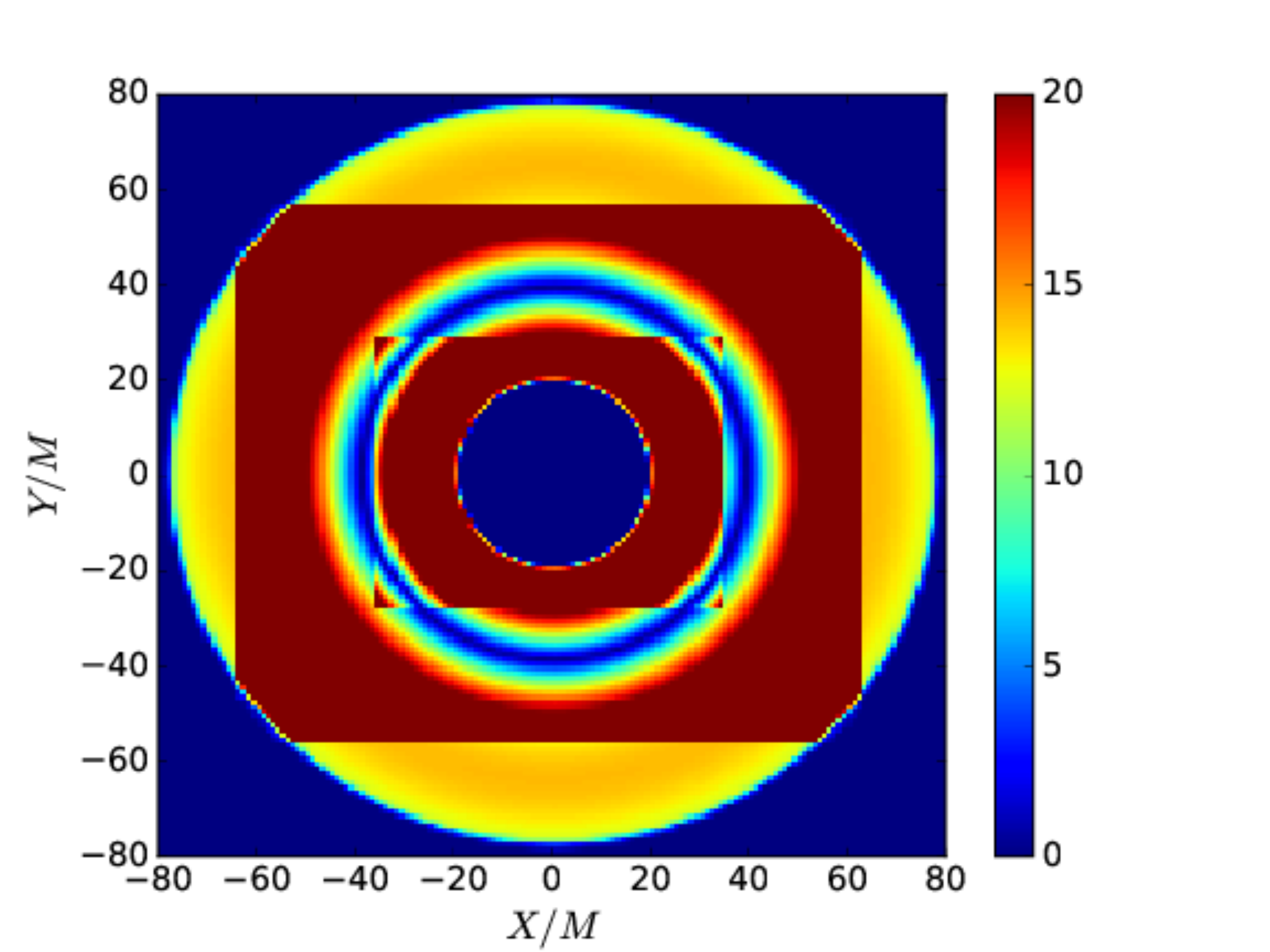}
\includegraphics[scale=0.29]{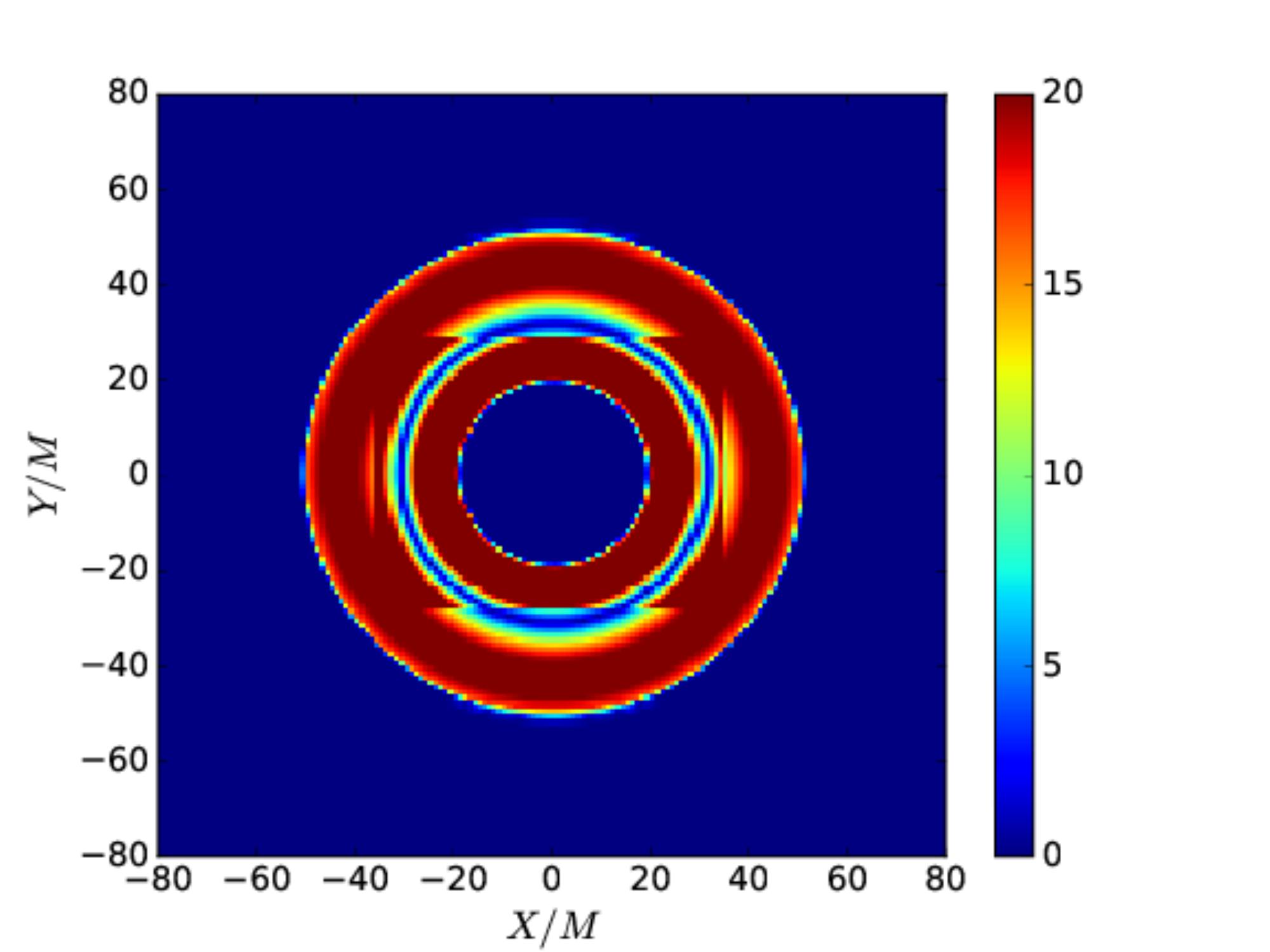}
\includegraphics[scale=0.29]{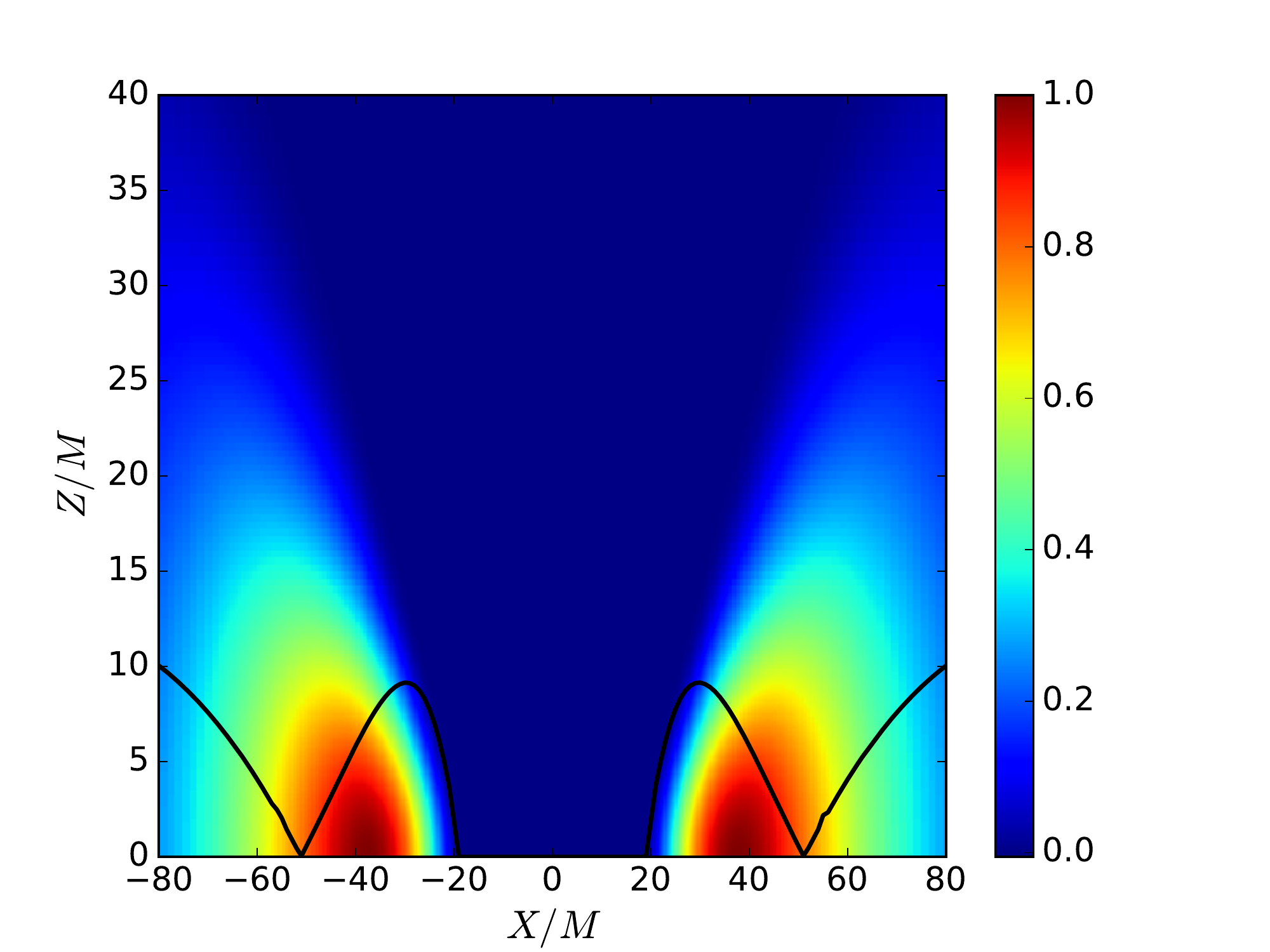}
\includegraphics[scale=0.29]{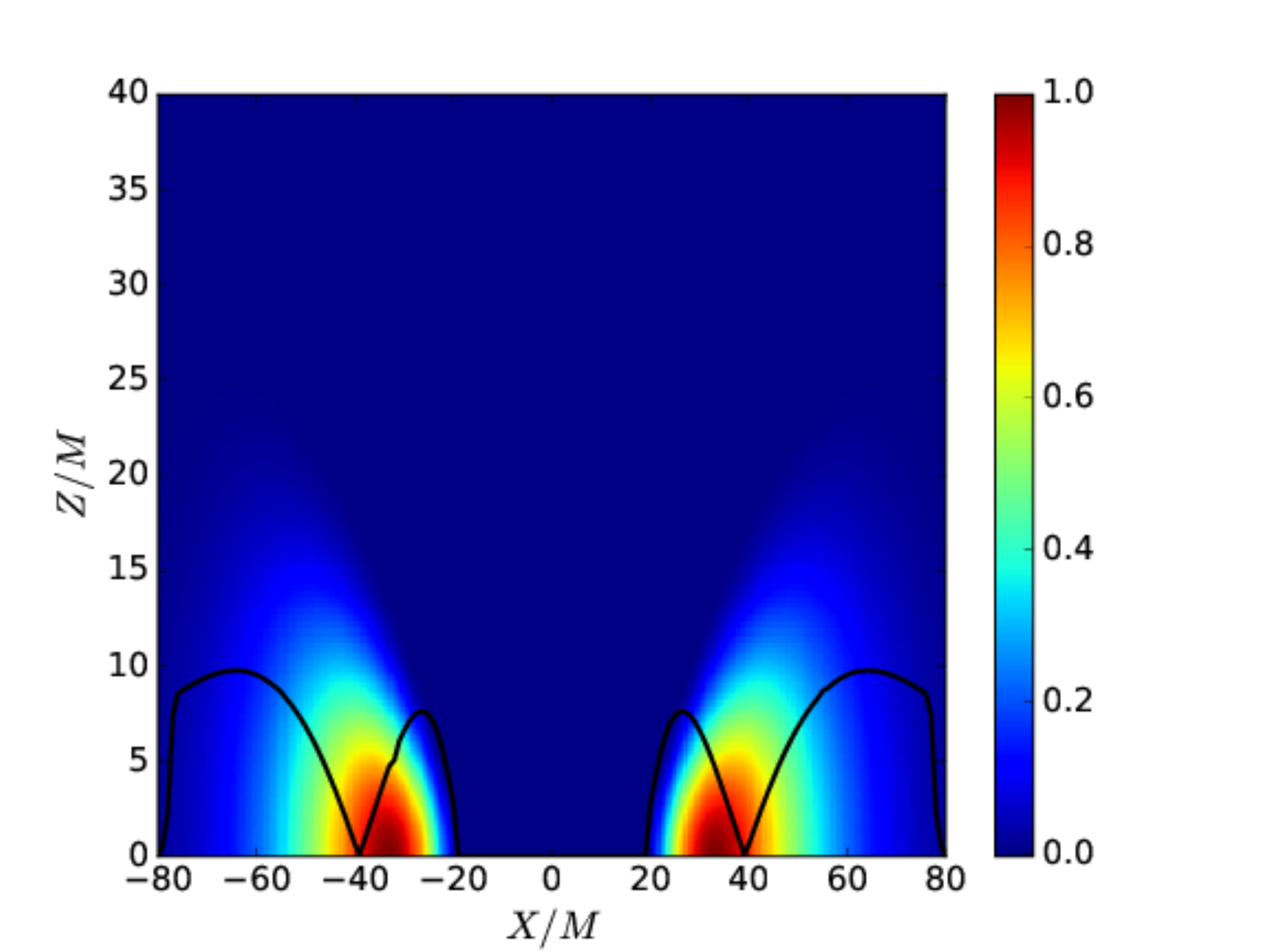}
\includegraphics[scale=0.29]{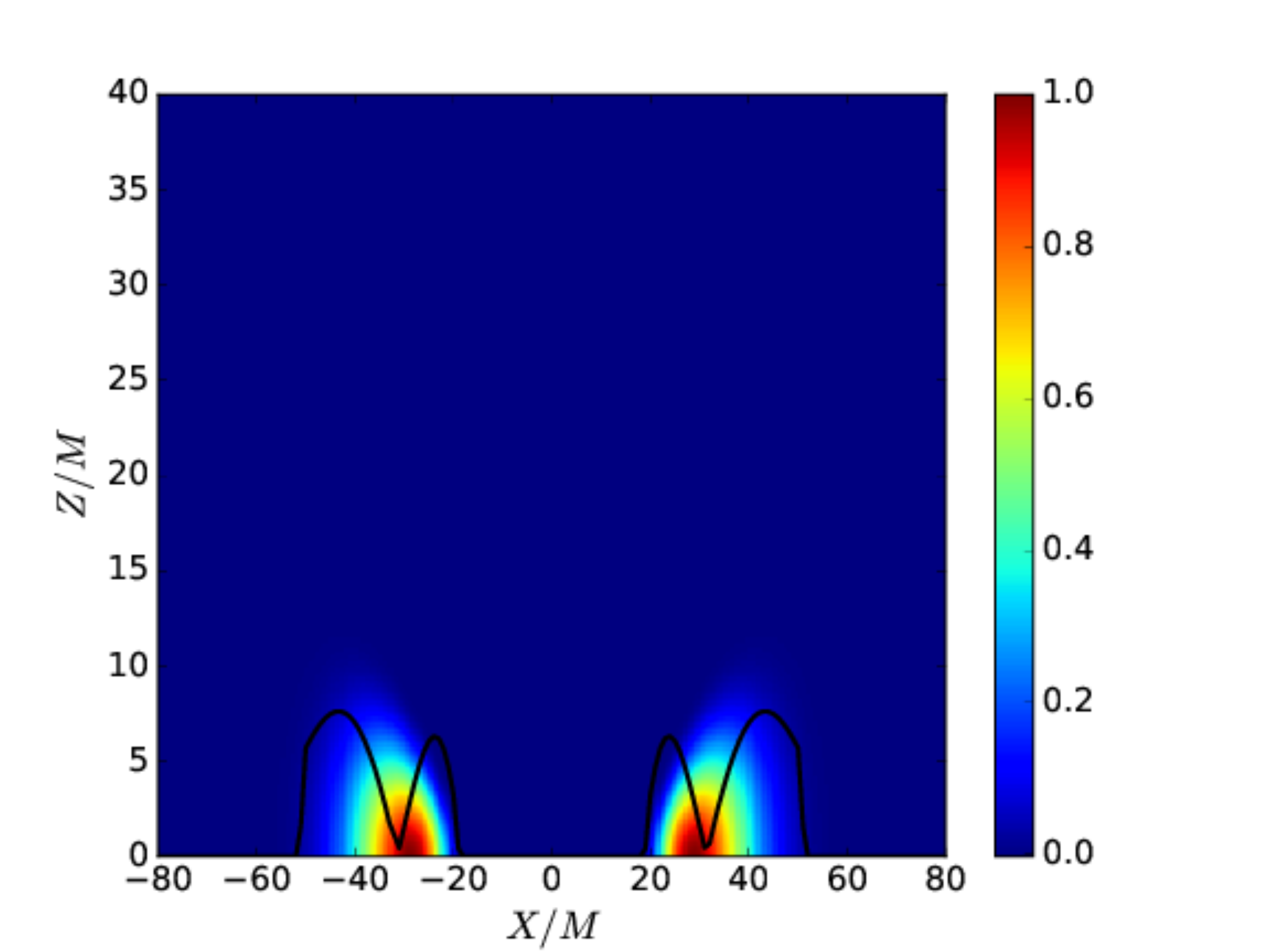}
\caption{Top row: Contours of the $\lambda_{\text{MRI}}$-quality
  factor $Q = \lambda_{\text{MRI}}/dx$ in the equatorial plane at
  $t=0$ for all cases.  Bottom row: Rest-mass density contours (color
  coded) on a meridional slice, and $\lambda_{\text{MRI}}/2$ (black
  solid line) at $t/M = 0$ for all cases. The left column plots
  correspond to case A, the middle column to case B and the right
  column to case C.  These plots demonstrate that with our grid
  choices we resolve the fastest growing MRI mode by $\gtrsim 10$
  points in the bulk of the disks (the blue ring stems from the
  extremely small values of $\lambda_{\text{MRI}}$ where the vertical
  component of the B-field changes sign). The plots also demonstrate
  that $\lambda_{\text{MRI}}/2$ mostly fits within the disks.
\label{fig:mri}}
\end{figure*}

As in~\cite{farris2}, we seed the initial disk with an initially
dynamically unimportant, purely poloidal magnetic field that is
entirely confined in the disk interior. The initial magnetic field is
generated by the following vector potential
\begin{align}
A_{i} &= \left(-\frac{y}{\varpi^{2}}{\delta^{x}}_{i} 
		+\frac{x}{\varpi^{2}}{\delta^{y}}_{i}\right)A_{\varphi} \\
A_{\varphi} &= A_{b}\varpi^{2}{\rm max}(P - P_{c},0)
\end{align}
where $\varpi^{2} = x^{2} + y^{2}$, and $A_{b}$, and $p_{c}$ are free
parameters. $A_b$ is determined by the maximum value of the
magnetic-to-gas-pressure ratio which we set to $P_{\rm mag}/P_{\rm
  gas} \approx 0.022, 0.028, 0.042$ for cases A, B, C,
respectively. We also choose the cutoff pressure $P_{c}$ to be 1\% of
the maximum pressure. The magnetic field renders the disk unstable to
the magnetorotational instability (MRI) that ultimately leads to the
development of MHD turbulence. The associated effective turbulent
viscosity allows angular momentum to be transported and accretion to
proceed. We impose three conditions that enable MRI to operate in our
disk models: (i) We choose disk configurations whose rotation profile
satisfies $\partial_{R}\Omega < 0$, where $\Omega=u^{\phi}/u^{t}$ is
the fluid angular velocity~\cite{duez2}. (ii) We resolve the
wavelength of the fastest growing MRI mode $\lambda_{\text{MRI}}$ by
$\gtrsim$ 10 gridpoints~\cite{shibata}. Using the initial disk data,
we plot the quality factor $Q\equiv\lambda_{\text{MRI}}/dx$, where
$dx$ is the local grid spacing, which jumps by a factor of 2 at
adaptive mesh refinement (AMR) boundaries.  The top panels of
Fig.~\ref{fig:mri} show plots of $Q$ in the equatorial $x-y$ plane for
all three cases, demonstrating that $\lambda_{\text{MRI}}/dx >10$. 
Note that the square patterns are a result of the AMR grids (see below).
(iii) The initial B-field is sufficiently weak --
$\lambda_{\text{MRI}} \lesssim 2H$, where $H$ is the disk scale
height, i.e., the wavelength of the fastest growing mode fits within
the disk. The bottom panels in Fig.~\ref{fig:mri} show meridional
xz-slices of the rest-mass density overlayed by a line plot showing
$\lambda_{\text{MRI}}/2$ as a function of $x$ for the three cases. For
the most part, $\lambda_{\text{MRI}}/2$ fits inside the
disk\footnote{$\lambda \lesssim 2H$ is not a requirement for resolving
  MRI, but an approximate requirement for the disk to be MRI unstable
  (see Eq. 2.32 of~\cite{Balbus:1991ay}). MRI is a weak field
  instability, so seeding the disk with a very strong B-field will
  render the disk MRI stable.}.

\begin{figure*}
\centering
\includegraphics[scale=0.218]{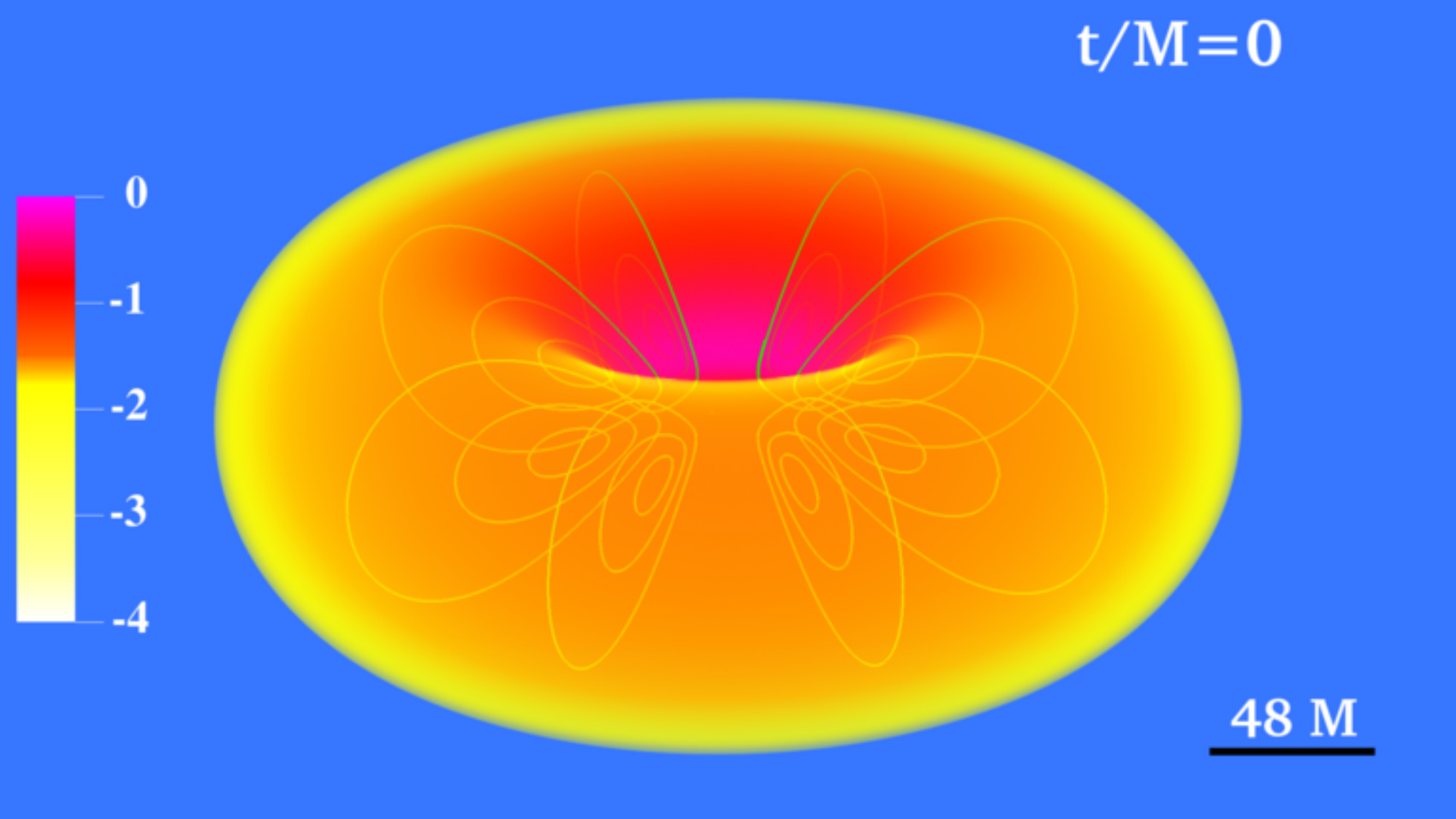}
\includegraphics[scale=0.218]{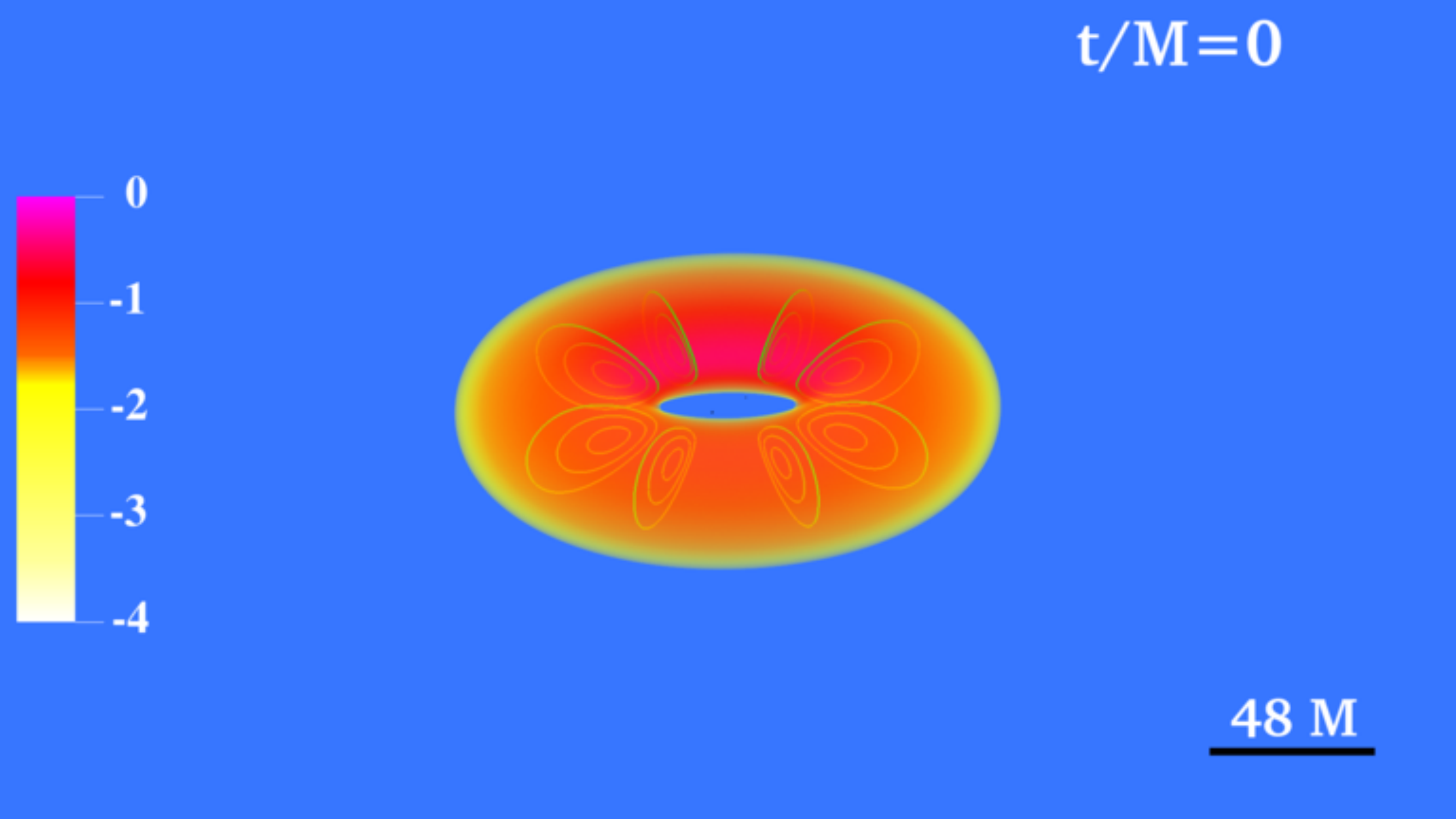}
\includegraphics[scale=0.218]{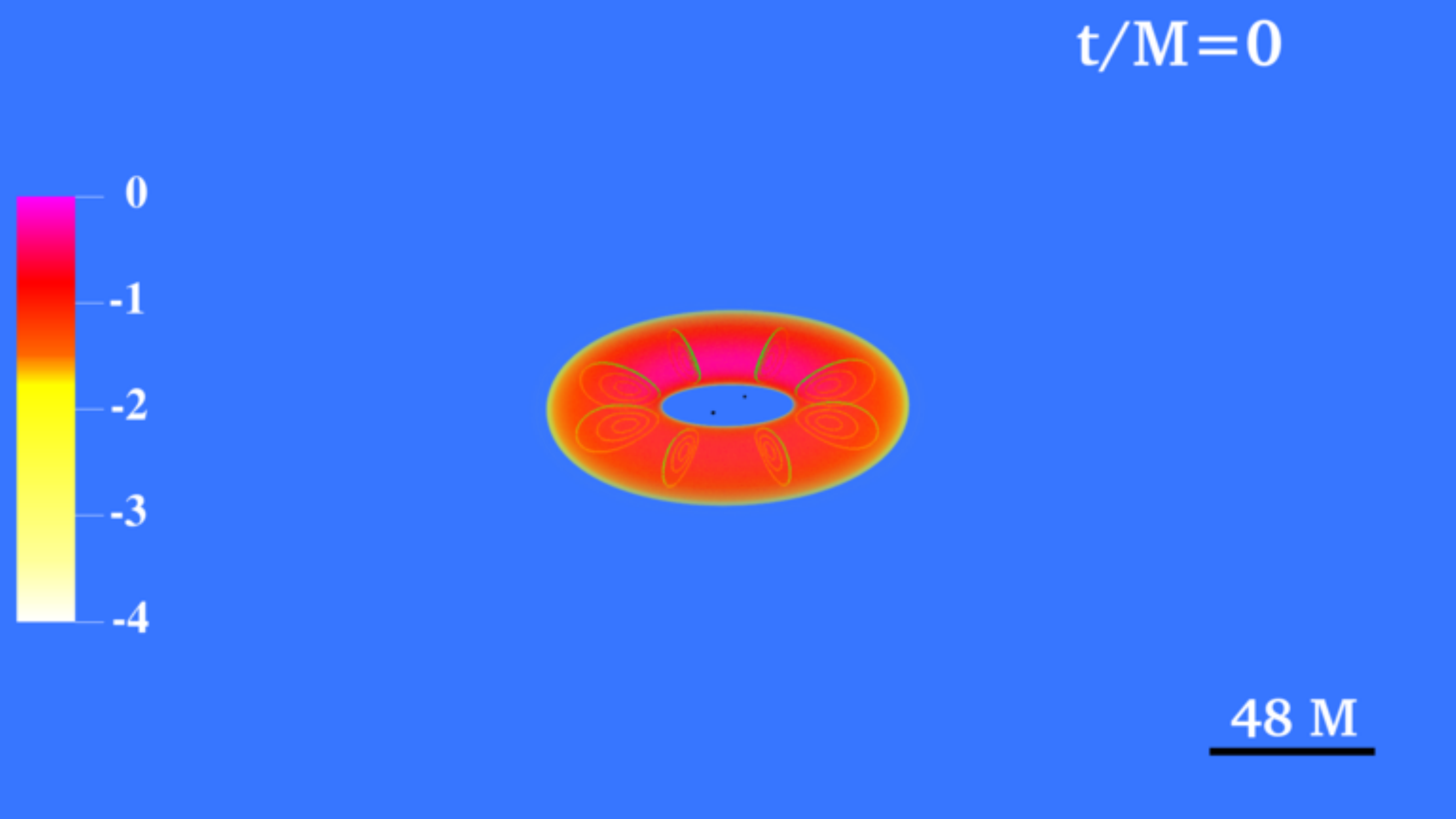}

\includegraphics[scale=0.218]{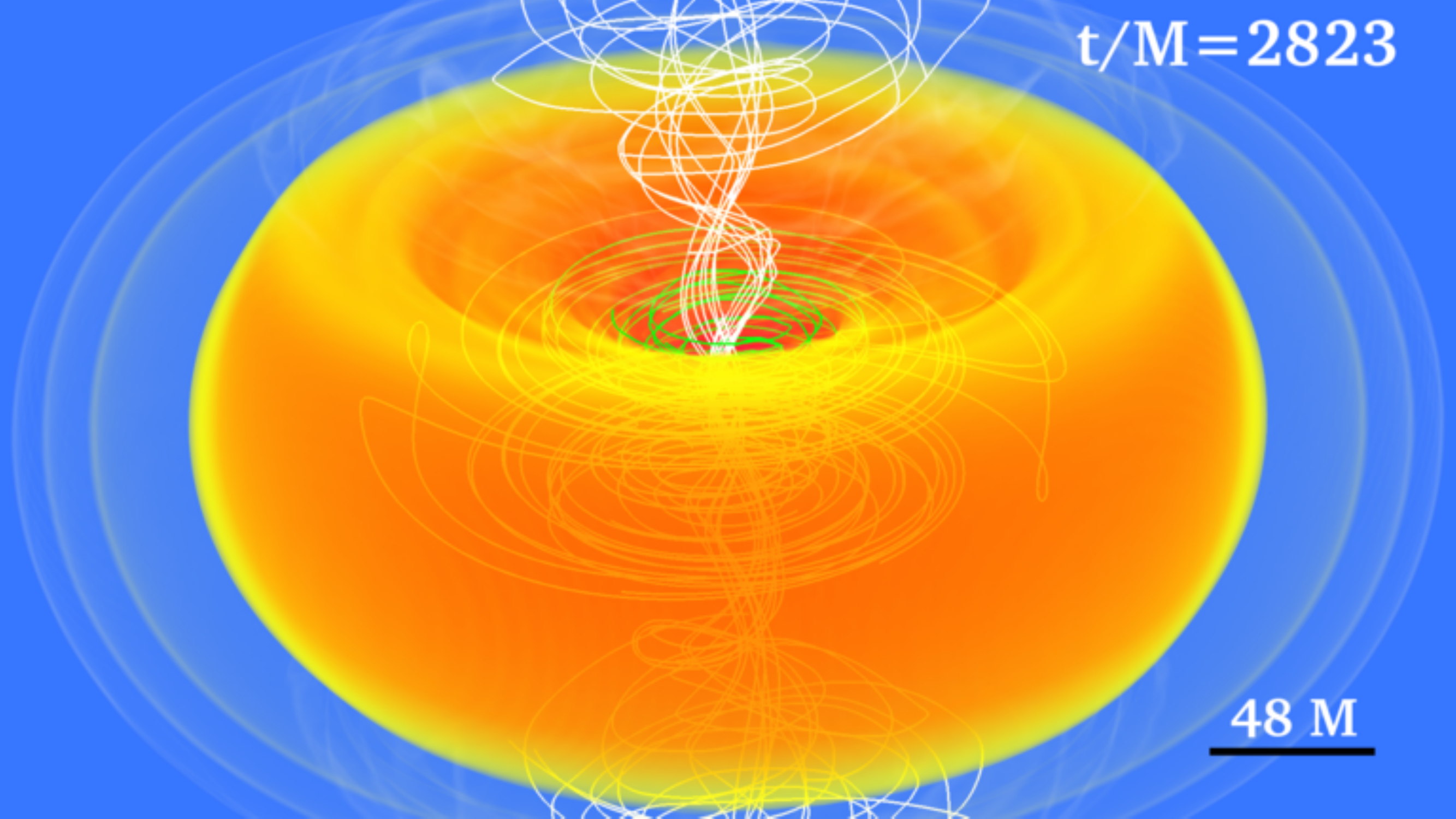}
\includegraphics[scale=0.218]{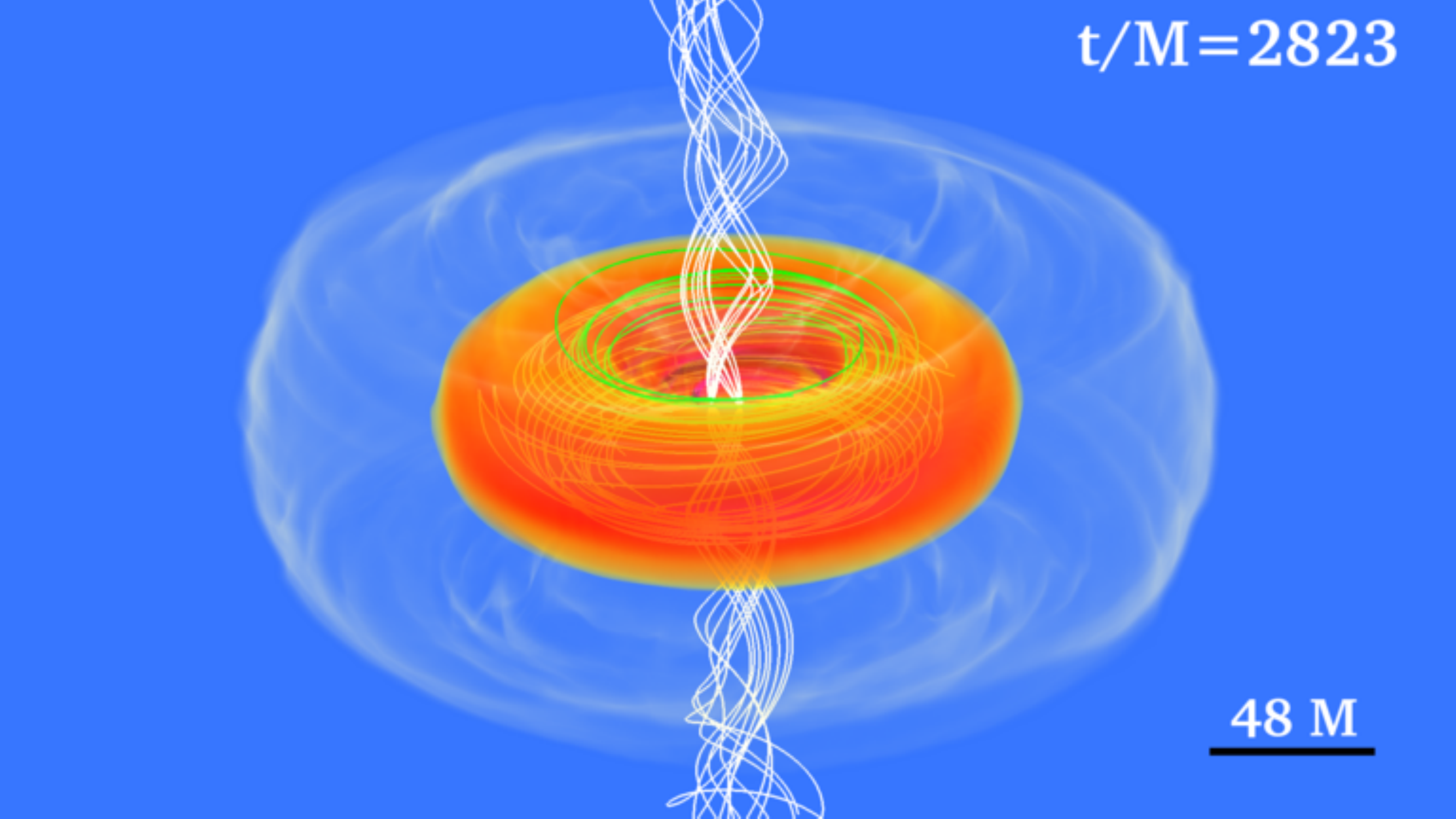}
\includegraphics[scale=0.218]{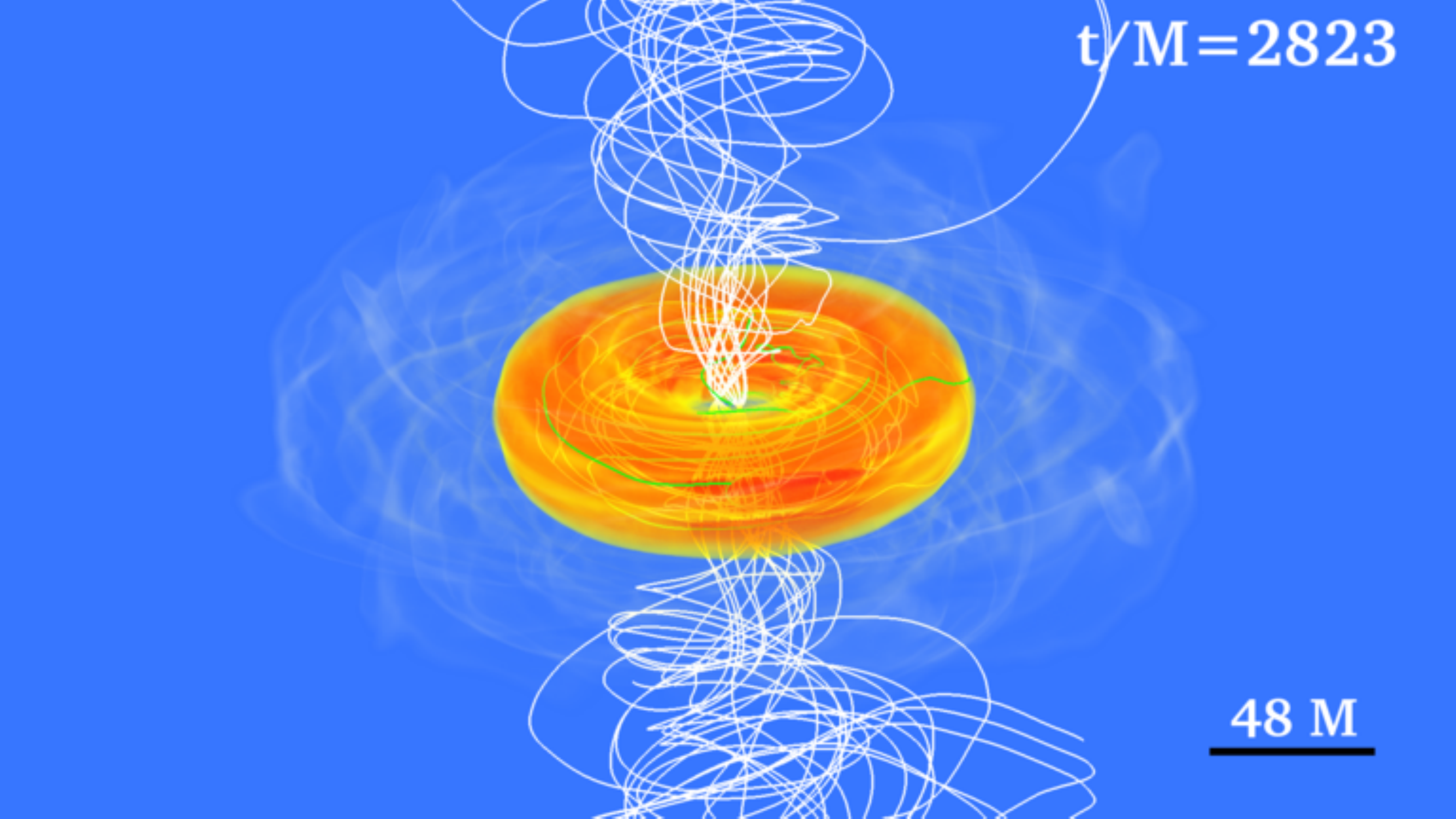}

\includegraphics[scale=0.218]{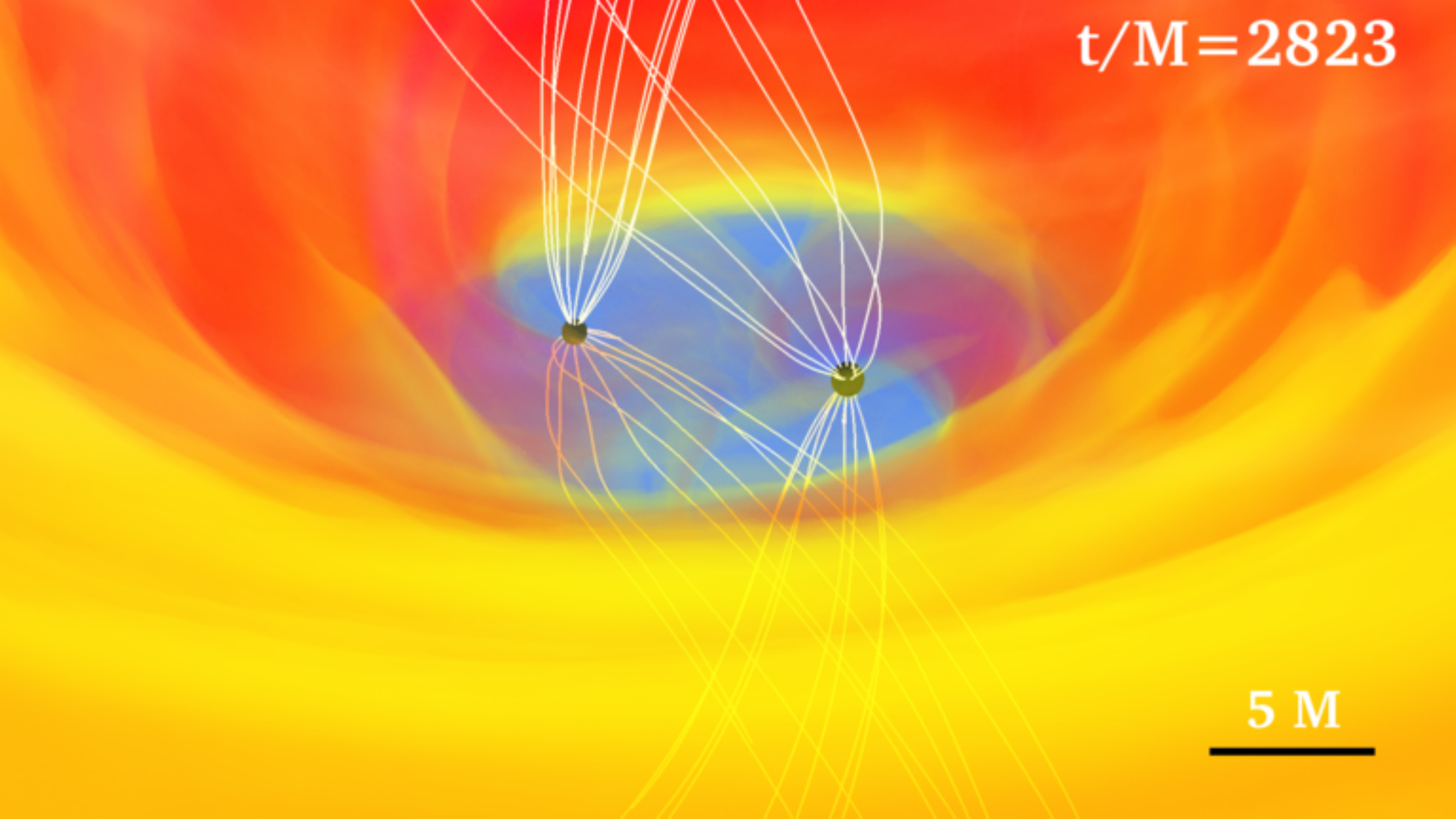}
\includegraphics[scale=0.218]{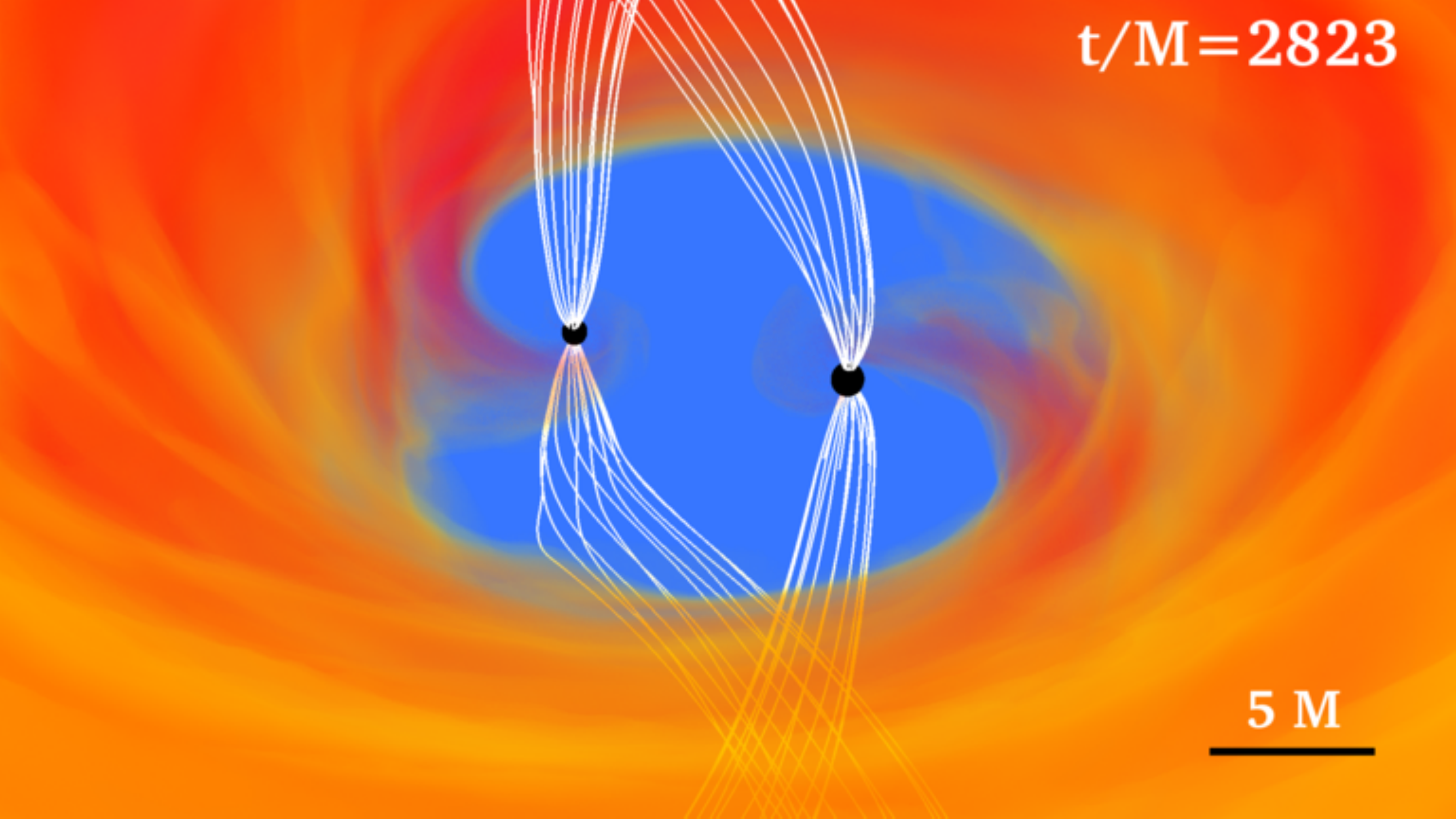}
\includegraphics[scale=0.218]{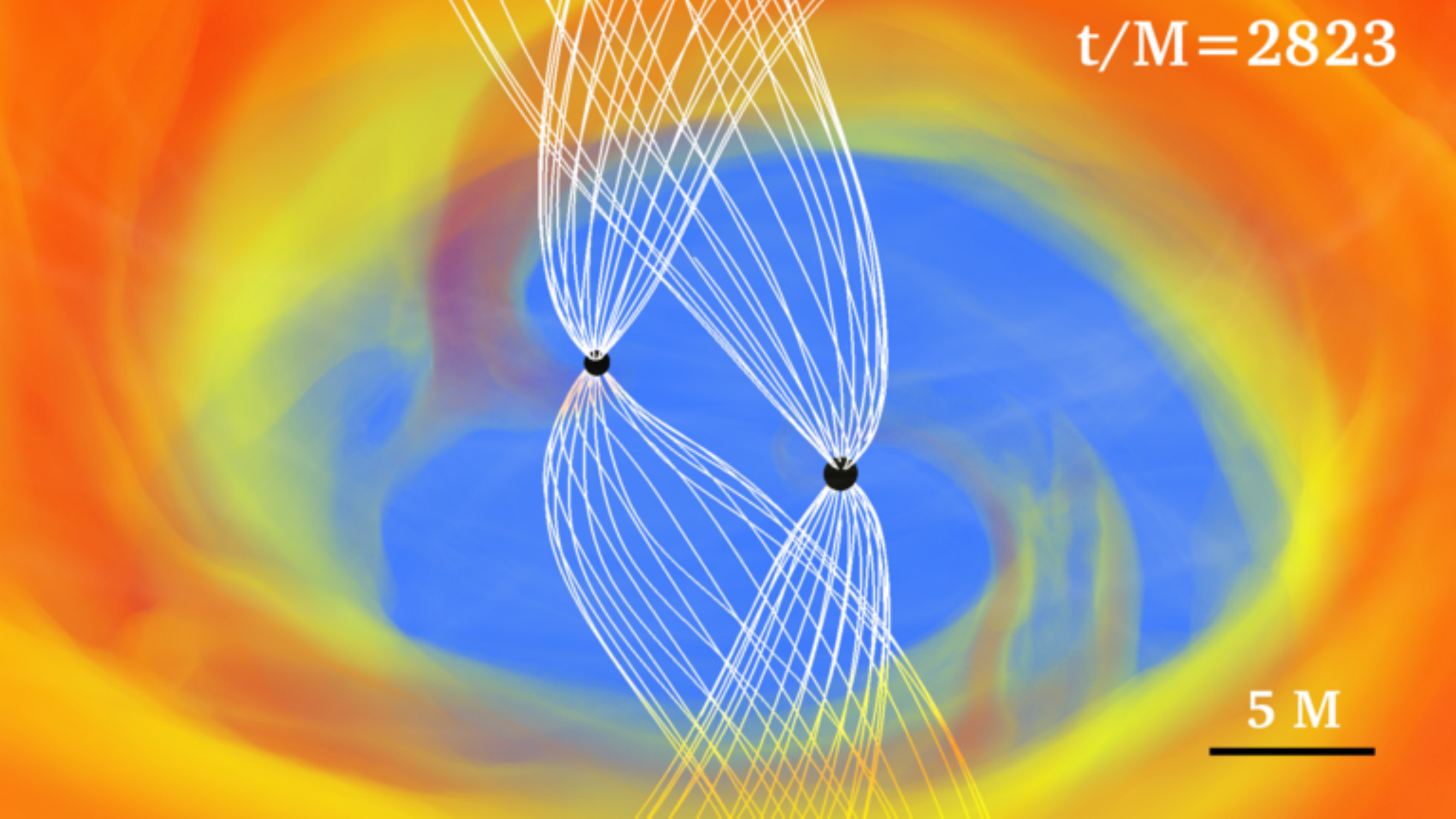}

\includegraphics[scale=0.218]{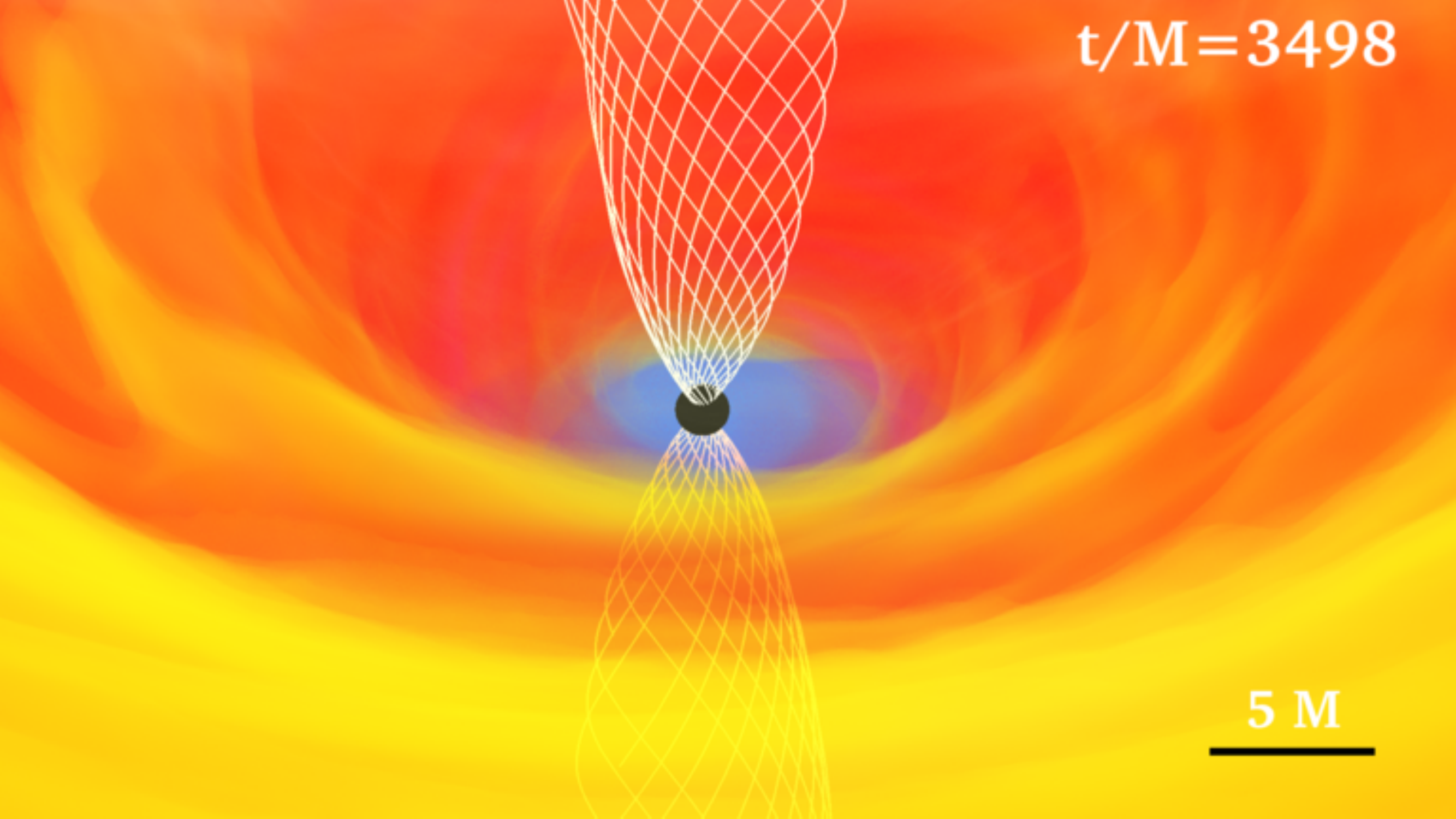}
\includegraphics[scale=0.218]{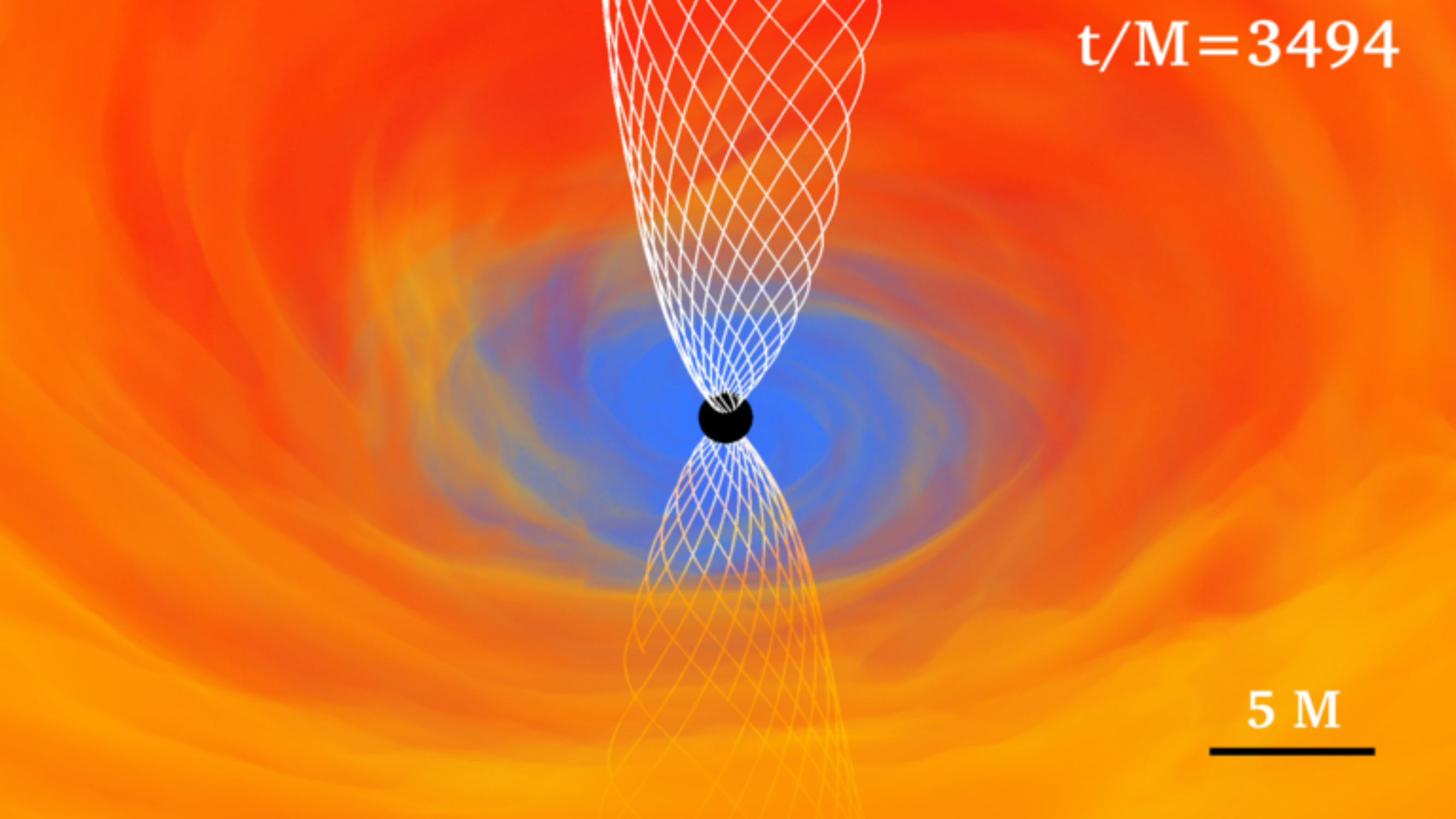}
\includegraphics[scale=0.218]{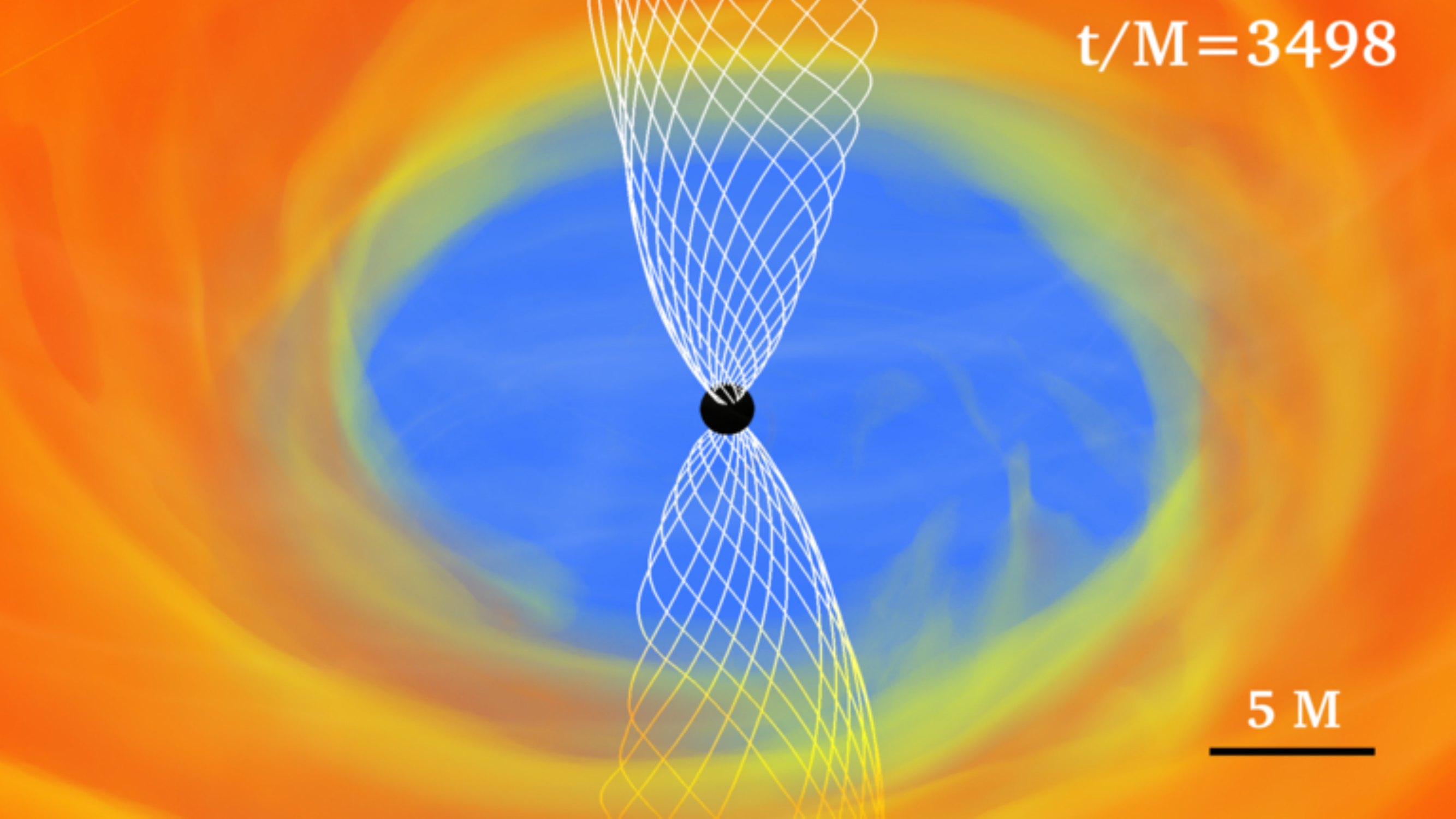}

\includegraphics[scale=0.218]{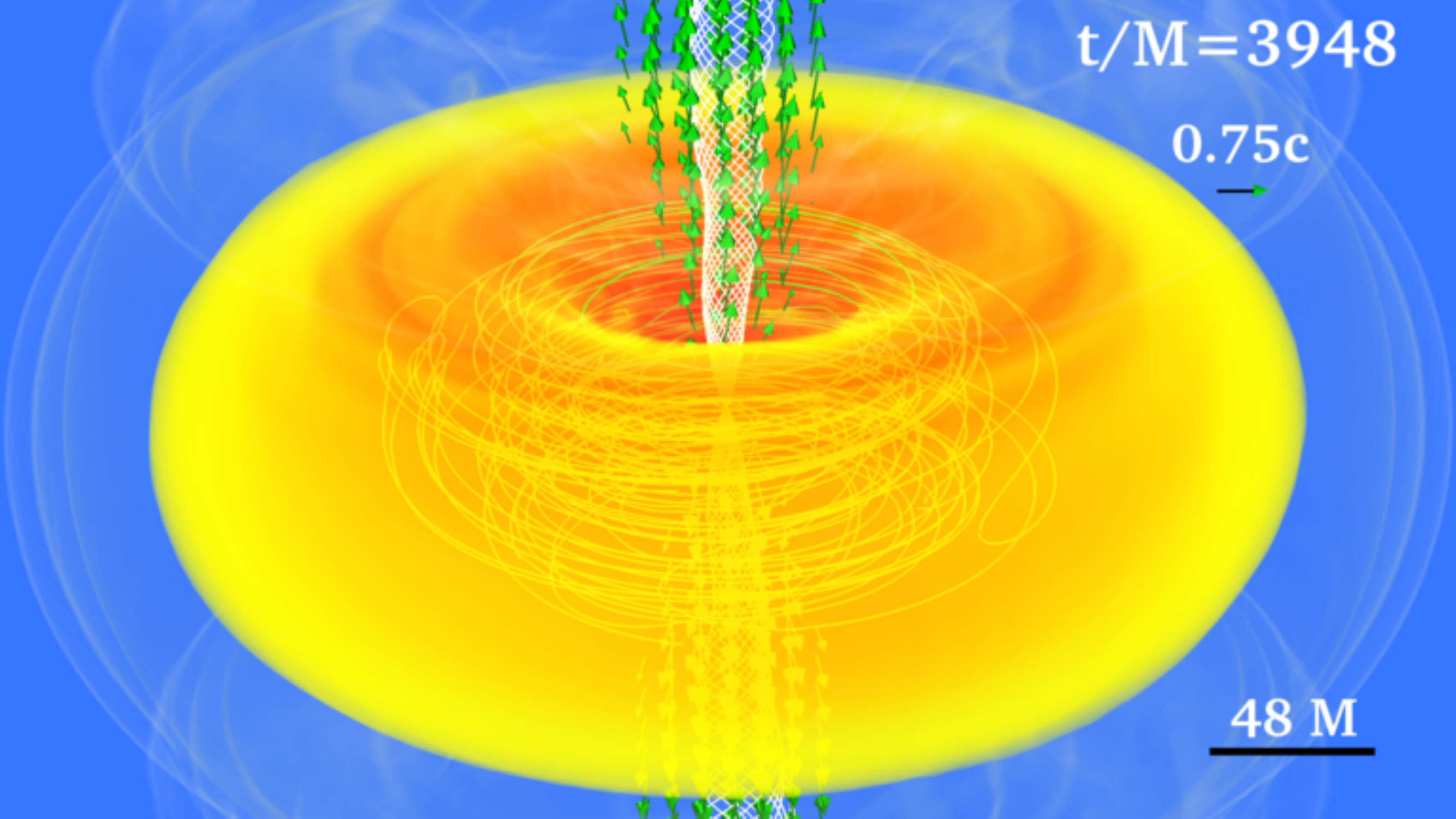}
\includegraphics[scale=0.218]{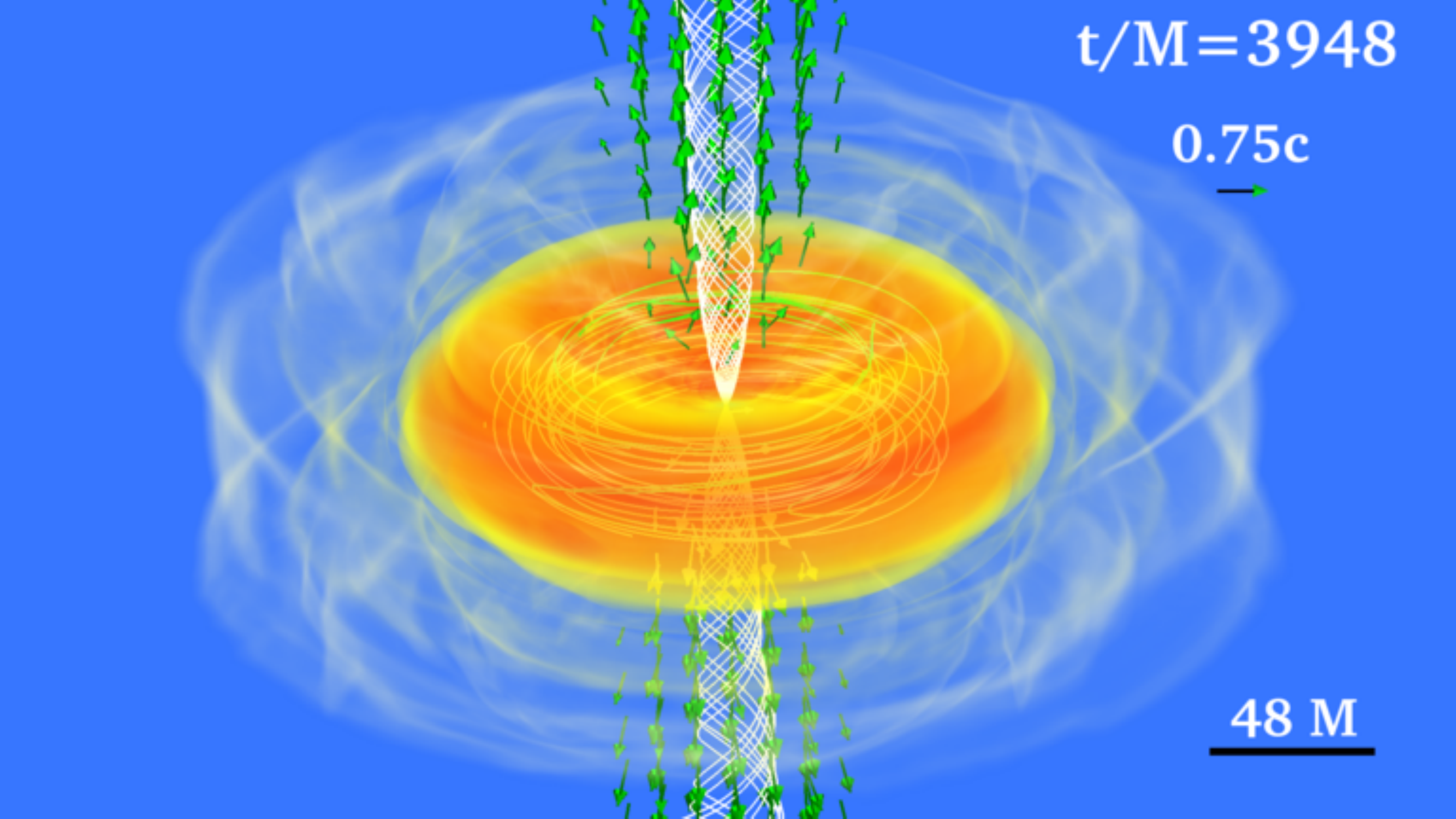}
\includegraphics[scale=0.218]{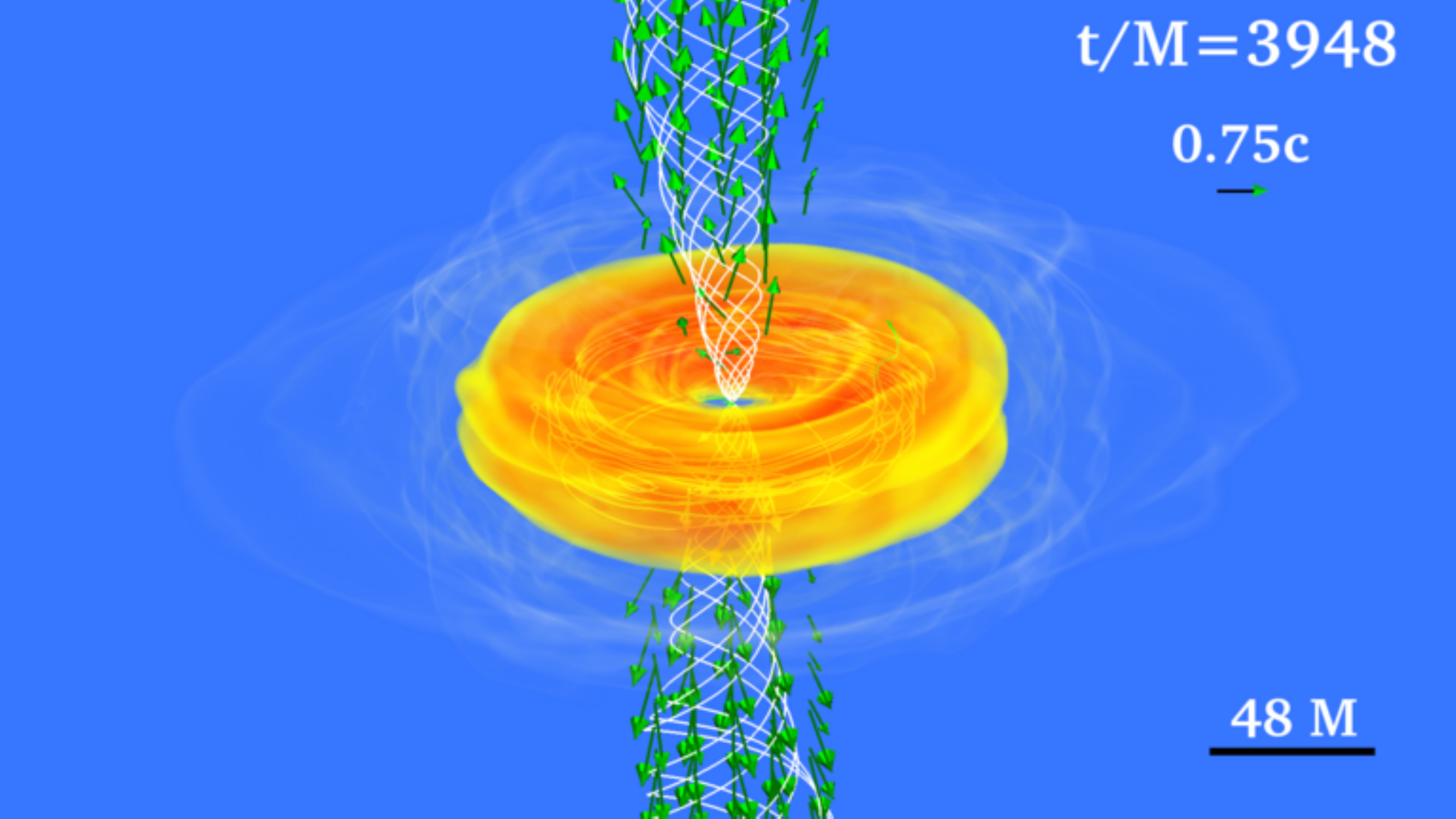}

\caption{Volume rendering of rest-mass density, normalized to its
  initial maximum value $\rho_{0, \text{max}}$ (see color coding),
  magnetic field lines (solid white curves), and velocity vectors
  (green arrows) at select times during the inspiral, merger and
  post-merger. Case A corresponds to the left column, case B to the
  middle column, and case C to the right column.
\label{fig:visit}}
\end{figure*}

\begin{figure}
\centering
\includegraphics[scale=0.32]{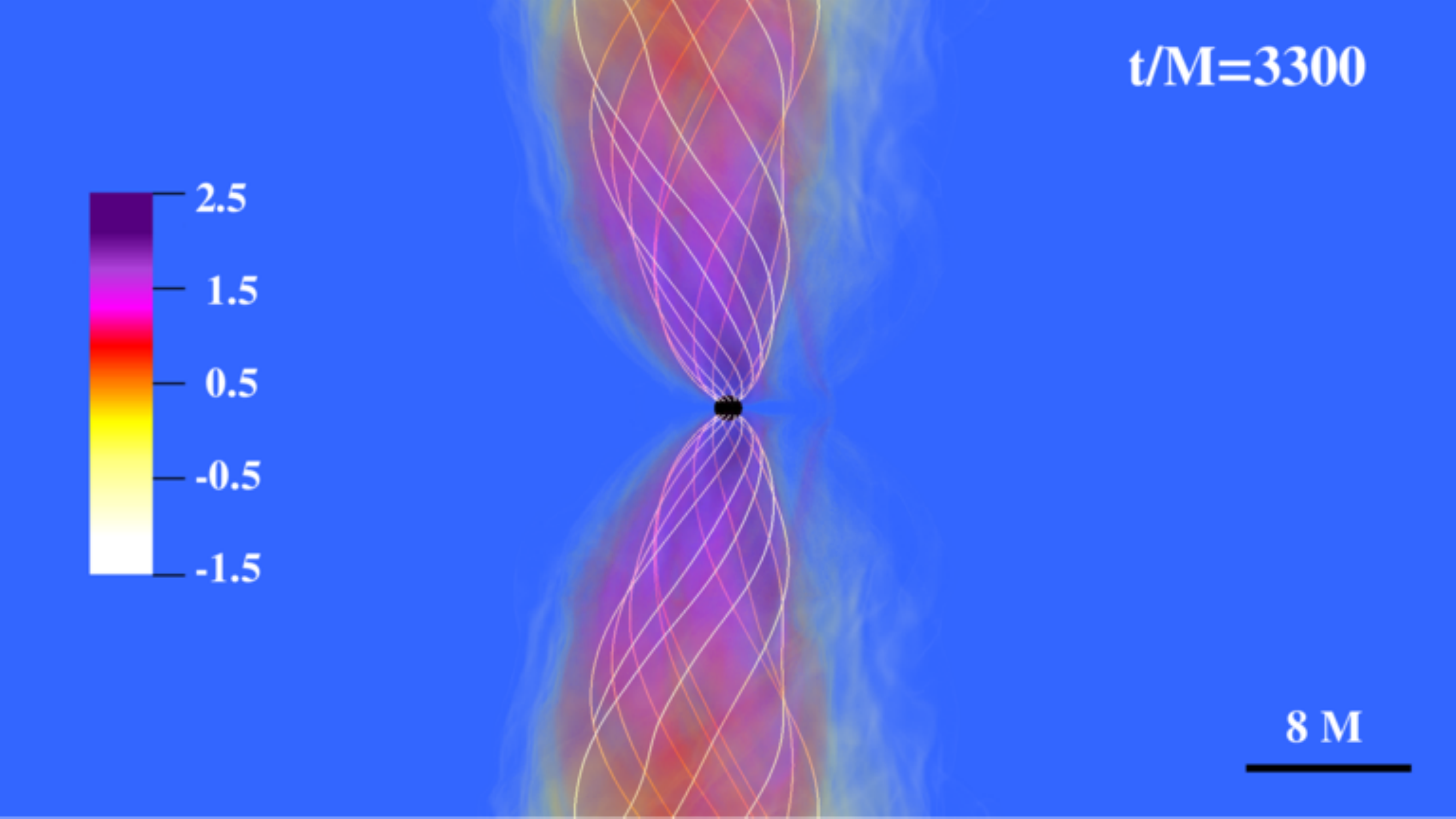}
\caption{Volume rendering of magnetic-to-rest-mass energy density
  $\log\left(b^{2}/2\rho_{0}\right)$ for case B. The white lines
  indicate the magnetic field lines corresponding to the B field
  measured by a normal observer.
\label{fig:b2rho}}
\end{figure}

\subsection{Dynamical evolution methods}

We use the Illinois GRMHD AMR code which is embedded in the
\texttt{Cactus/Carpet} infrastructure~\cite{cactus, carpet} and has
been developed by the Illinois relativity group~\cite{duez, etienne,
  etienne2}. This code is the basis of its publicly available
counterpart embedded in the Einstein Toolkit~\cite{Etienne:2015cea}.
Our code has been extensively tested (including with resolution
studies) and used in the past to study numerous systems involving
compact objects with and without magnetic fields \cite{etienne4,
  etienne3, liu2, vasilis2, vasilis3} including black hole binaries in
magnetized accretion disks~\cite{gold, gold2}.

For the metric evolution, the code solves the equations of the
Baumgarte-Shapiro-Shibata-Nakamura (BSSN) formulation of
GR~\cite{Shibata95,BS} coupled to the moving-puncture gauge
conditions~\cite{BakCenCho05,CamLouMar05} with the equation for the
shift vector cast in first-order form (see
e.g.~\cite{HinBuoBoy13}). The shift vector parameter $\eta$ is set
to~$\eta = 1.375/M$.

For the matter and magnetic field, the code solves the equations of
ideal GRMHD in flux-conservative form (see Eqs. 27-29 in
\cite{etienne}) employing a high-resolution-shock-capturing scheme. To
enforce the zero-divergence constraint of the magnetic field, we solve
the magnetic induction equation using a vector potential formulation
(see Eq. 9 in \cite{etienne2}). As our EM gauge choice, we use the
generalized Lorenz gauge condition developed in \cite{farris2}.  This
EM gauge choice avoids the development of spurious magnetic fields
that arise across AMR levels (see~\cite{EtVpas12} for more details).
We set the generalized Lorenz gauge damping parameter to $\xi = 7/M$.

To compare with our previous calculations~\cite{gold,gold2} and to
reduce the computational overhead associated with the evolutions,
equatorial symmetry is imposed in all cases. Equatorial symmetry does
not allow plasma to cross the equatorial
plane~\cite{etienne4,Yan2012,Roberts2017} and hence limits the
development of some modes. We will explore the impact of reflection
symmetry on such circumbinary accretion flows in a future
investigation. The computational mesh consists of three sets of nested
AMR grids, one centered on each BH and one on the binary center of
mass (with $9$ levels of refinement in each set). The outer boundary
is at $384$M in all cases. The coarsest grid spacing is $\Delta x_{\rm
  max}=6.0$M. The grid spacing of all other levels is $\Delta x_{\rm
  max}/2^{n-1},\ n = 1, 2,\ldots$, where $n$ is the level number such
that $n = 1$ corresponds to the coarsest level. The half length of
each AMR level box is $384{\rm M}/2^{n-1}$ for $n =1,\ldots 5$, and
$528{\rm M}/2^{n}$, for $n =6,\ldots,9$.


\section{Results}
\label{sec:results}

The basic evolution of the BHBH-disk system has previously been
described in~\cite{farris2,gold,gold2}, where it was found that when
keeping the binary at fixed orbital separation of $10M$ for a disk
with initial inner radius $R_{\rm in} = 20M$, the accretion rate
begins to settle at $t\simeq 2000M$, which signals that the inner disk
edge has relaxed. This result motivates in part our choice of initial
BHBH orbital separation of $13.8$M, because within a time of 2000M the
binary undergoes about 10 orbits and then reaches the orbital
separation of $\sim 10M$, at which point the inspiral in~\cite{gold2}
began, once the disk was allowed to relax. The fact that the initial
orbital separation in our current simulations is larger than in our
earlier studies implies that the accretion rate is expected to settle
even earlier for the same disk model, because the black holes are
closer to the disk and hence it is easier for them to tidally strip
matter from the inner disk edge. Our simulations confirm this
expectation, where for the same initial disk model as
in~\cite{farris2,gold} the accretion rate begins to settle at $t \sim
1500M$.

\subsubsection{Evolution of the matter and magnetic fields}

Figure~\ref{fig:visit} shows 3D volume snapshots of the density at
selected epochs during the evolution. Soon after the evolution begins
we observe the formation of high-density, spiral accretion streams
that attach onto the BHs during the entire inspiral and through merger
(see the middle row in Fig.~\ref{fig:visit}).  In all cases we also
observe the emergence of a mildly relativistic, collimated outflow
from each black hole. These outflows merge at large heights to form a
single helical magnetic-field, incipient jet flow (second row of
Fig.~\ref{fig:visit}). As the binary inspirals the incipient twin jets
merge more tightly and the inner disk edge flows inward, both helping
to collimate the outflow further. After merger a single incipient jet
onto the remnant black hole forms~(4th and 5th rows in
Fig.~\ref{fig:visit}).

\begin{figure*}[ht]
\centering
\includegraphics[scale=0.59]{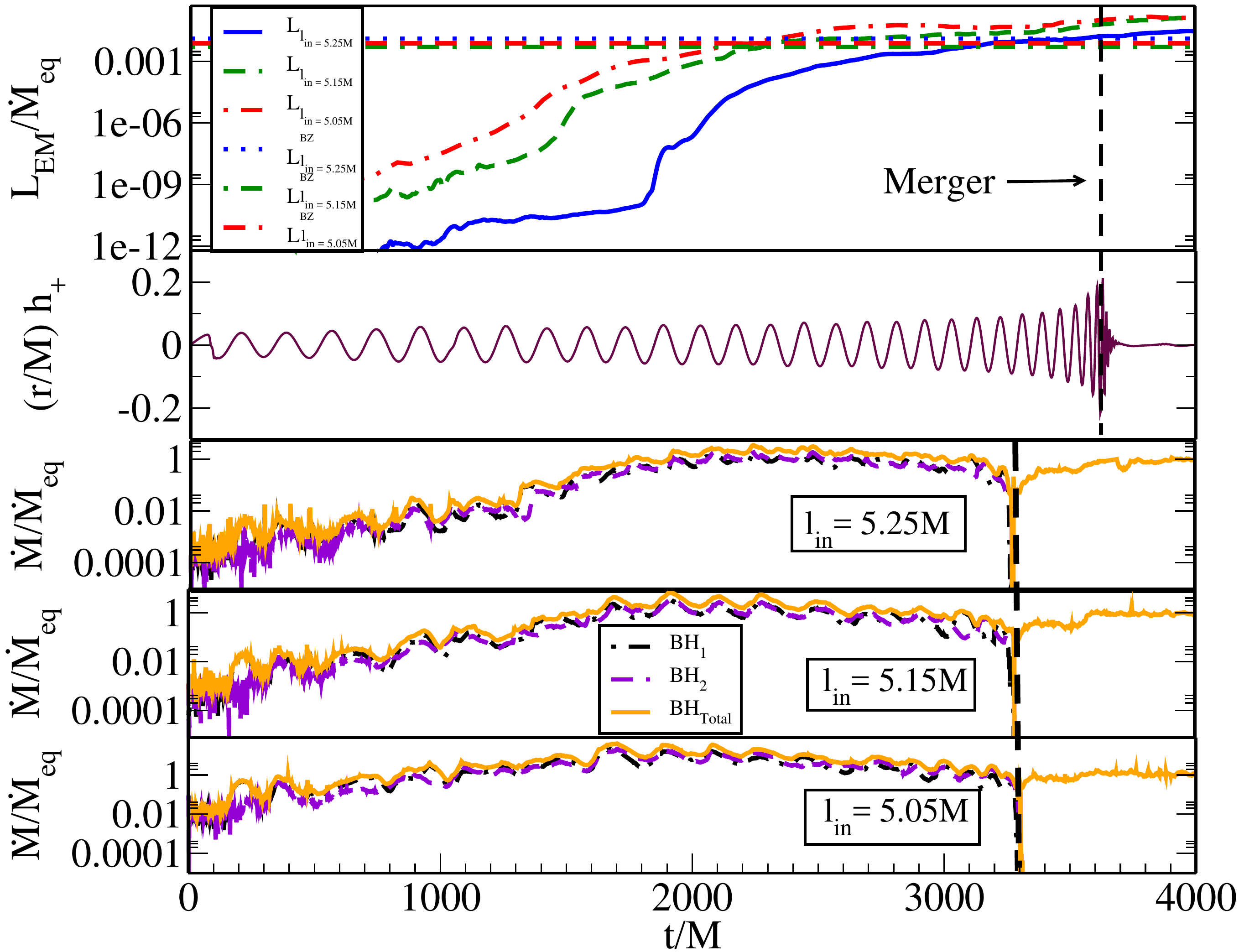}
\caption{The Poynting luminosity $L_{\rm EM}$ and accretion rate
  $\dot{M}$ vs. time for all cases and the strain of the ``plus''
  polarization of the gravitational waveform $h_{+}$. The GWs are
  computed as described in~\cite{Etienne:2011ea}. The dotted-dashed
  lines on the top panel indicate the expected Blandford-Znajek
  luminosity from a single BH with the same mass, spin and
  quasistationary polar magnetic field-strength as found for each
  case. The normalization $\dot M_{\rm eq}$ in each panel is the
  time-averaged accretion rate onto the post-merger remnant black hole
  over the last 500M of the evolution. The displacements in the dashed
  vertical merger line in the luminosity and GW plots with respect to
  the merger lines in the accretion plots accounts for the light
  travel time between the BH and the radiation extraction radii.
\label{fig:1dplots}}
\end{figure*}

\subsubsection{Accretion rates}

We compute the accretion rate through the black hole apparent horizons
via Eq. (A11) of~\cite{Farris:2009mt}. Figure~\ref{fig:1dplots} shows
the accretion rate of all cases as a function of time, along with the
outgoing Poynting luminosity. Also shown is the gravitational
waveform. The accretion rate reaches a quasisteady state at $t/M \sim
1500$ for all cases, and the average accretion rate remains
approximately constant for the next $1000$M of evolution, consistent
with the system being in the ``predecoupling'' phase, as
anticipated. During this phase a Fourier transform indicates that the
accretion rate exhibits a periodicity at periods equal to $202$M,
$210$M and $202$M for cases A, B and C, respectively. These periods
equal $\simeq 0.8P$, where $P$ is the average orbital period, $P\simeq
250$M, which we determine from the GWs during this predecoupling
epoch. Notice that the dominant period in the accretion rate being
smaller than the binary orbital period is consistent with our findings
in~\cite{gold}.  The $\sim 0.8P$ accretion rate periodicity has been
associated with a ``lump'' feature in earlier
work~\cite{Noble2012,Roedig2012}. However, we have reported accretion
periodicities at $0.7-0.8P$ in earlier work of ours~\cite{gold} even
in cases where a lump was absent. Thus, there may be more than one
mechanism that can explain this periodicity. We plan to address this
point with targetted simulations in a future work.  The fact that for
different disk models the accretion rate modulation occurs at
approximately the same period indicates that this signature is not
sensitive to the adopted disk models.  At $t\sim2500M$, the inspiral
of the BHs begins to speed up, and the accretion rate begins to drop,
indicating the onset of the ``postdecoupling'' phase. At the time of
merger the total accretion rate drops, but soon after merger, the
low-density hollow begins to fill up, increasing the accretion rate
after merger.

Following merger, for the thinnest disk case during the time evolved
the accretion rate achieves a steady-state value at $\sim 3\%$ of the
maximum accretion rate during the evolution. For the middle disk the
post-merger accretion rate is $\sim 14\%$ of the maximum accretion
rate achieved during evolution; for the thicker disk this value is
$\sim 30\%$.  A careful resolution study is necessary to assess the
significance of this result, which we defer to a future investigation.

\subsubsection{Outflows and jets}

In all three cases we find that at $t\sim 2500$M after the start of the
simulations, collimated and magnetically dominated outflows emerge in
the disk funnel independently of the disk model. The timescale over
which these incipient jets emerge is consistent with those we reported
in~\cite{farris2,gold,gold2}. The incipient jets persist through
inspiral, merger and post-merger, as seen in Fig.~\ref{fig:visit}.
During merger we find that the magnetization in the funnel grows and
after the merger the magnetization reaches $b^{2}/2\rho_{0} \gtrsim
100$, indicating that the jets are magnetically powered. Here
$b=B/4\pi$ with $B$ the magnetic field measured by a comoving
observer, and $\rho_{0}$ is the rest-mass density. The region above
and below the BH remnant is nearly force-free -- a prerequisite for
the Blandford-Znajek (BZ) mechanism~\cite{bz}. A representative plot
of $b^{2}/2\rho_{0}$ and the magnetic field lines in the funnel is
shown in Fig.~\ref{fig:b2rho}. Associated with these outflows is a
large scale outgoing Poynting flux.

We calculate the outgoing Poynting luminosity on the surface of
coordinate spheres $S$ as $L_{(EM)} = \oint_{S}T_{0,\, (EM)}^{\,\,\,
  r}dS$, where $T_{\mu,\, (EM)}^{\,\,\, \nu}$ is the EM stress-energy
tensor. $L_{(EM)}$ as a function of time is shown in
Fig.~\ref{fig:1dplots} for all cases. Following binary merger, the
luminosity ramps up to a nearly constant value in all cases for the
duration of the simulation.  After merger, we find that the
``efficiency'' $\epsilon_{\rm EM} \equiv L_{\rm EM}/\dot{M}_{\rm eq}$
($\dot M_{\rm eq}$ is the time-averaged accretion rate onto the
post-merger remnant BH over the last 500M of the evolution) increases
as the disk thickness decreases. However, we note that since our goal
here is to test the robustness of global BHBH-disk features across
different disk models, we did not prepare initial disks with the same
initial amount of poloidal magnetic flux, which has been found to be
one of the major determining factors for the efficiency of the
outgoing Poynting luminosity in MHD BH accretion disks (see
e.g.~\cite{Beckwith:2007sr,McKinney:2012vh}).

In~\cite{gold2} we discovered that following merger there is a boost
in the jet power by a factor 3-4 with time delay in units of M between
merger and the boost that depends on the mass ratio. In particular, we
found that the more unequal mass BHBHs, the shorter the time delay in
units of M, which ranges between $\sim 350$M (for $q=4$) and $\sim
1000$M (for $q=1$). Here, we discover that the boost in the Poynting
luminosity and the existence of a time delay between merger and the
boost is robust among the different disk models we consider in this
work. In our study the time delay is $\sim 600$M for cases A and B,
and $\sim 300$M for case C, suggesting that for fixed mass ratio and
thick disks, the time delay between merger and boost in EM luminosity
exhibits only modest dependence on the disk thickness.

After the boost and the outgoing Poynting luminosity achieve a
quasiequilibrium, the Poynting luminosity is comparable to the EM
power expected from the BZ effect:
\begin{equation}
  L_{\rm BZ}
  \sim10^{48}\left[\frac{(a/M_{\rm BH})}{0.68}\right]^{2}\left(\frac{M_{\rm BH}}{62M_{\odot}}\right)^{2}\left(\frac{B_{p}}{4
  \times 10^{12}\rm G}\right)^{2}\rm erg s^{-1}
\end{equation}
(see Eq. 4.50 in \cite{thorne}). We plot $L_{\rm BZ}$ normalized to
the post-merger accretion rate averaged over the last 500M in the top
panel of Fig.~\ref{fig:1dplots}, taking the remnant BH dimensionless
spin and mass values from our simulations, while adopting for the
magnetic field above the BH poles $B_p$, its time-averaged value over
the last 500M of evolution.

As mentioned earlier, above the remnant BH poles the funnel is nearly
force-free ($B^{2}/8\pi\rho_{0} \gg 1)$). The BZ solution has been
found to approximately describe the force-free regions in the funnel
of magnetized, geometrically thick disks accreting onto single,
spinning BHs \cite{mckinney,DeVilliers:2004zz}, thus, we can attribute
this luminosity and accompanying incipient jet from the remnant
spinning BH-disk system to the BZ effect. During the inspiral a
``kinetic'' or ``orbital'' BZ effect can account for the outgoing EM
energy~\cite{Palenzuela:2010nf}.


\section{Astrophysical implications}
\label{sec:astro}

In this section we discuss the implications of our results to
astrophysical systems of interest, ranging from black hole binaries
relevant to LIGO to supermassive black hole binaries that likely
reside at the centers of AGNs and quasars.

\subsection{LIGO GW150914}

\begin{table*}[th] 
  \caption{Table for the maximum rest-mass density $\rho_{0}^{max}$,
    disk rest-mass to BHBH mass ratio $M_{\text{disk}}/M_{\text{BH}}$,
    post-merger, time-averaged accretion rate $\dot{M}_{\rm eq}$ over
    the last 500M of the evolution, maximum equivalent isotropic
    luminosity $L_{\rm iso}$ after merger, the magnetic-field strength
    measured by an observer comoving with the plasma averaged over a
    cube of side equal to the remnant coordinate BH apparent horizon
    radius $r_{\rm AH}$ and located immediately above the remnant BH
    at the end of the simulation, and the time $\tau_{\text{delay}}$
    between merger and peak luminosity.
    \label{tab:t3}}
  \begin{center}
    \begin{tabular}{ccccccc} 
      \hline
      \multicolumn{7}{c}{Simulation Results (for GW150914)}\\
      \hline
      \hline
      Case & $\rho_{0}^{\text{max}}$ (g cm$^{-3})$ & $M_{\text{disk}}/M_{\text{BH}}$ & 
      $\dot{M}_{\rm eq}$ ($M_{\odot}$ s$^{-1}$) & $L_{\rm iso}$ (erg s$^{-1}$) & 
      $B_{p}$ (G)   & $\tau_{\text{delay}}$ (s)\\
      \hline
      A & $2.4\times 10^{5}$ & $1.1\times 10^{-3}$ & $3.3\times 10^{-4}$ &
      $1.8\times10^{49}$  & $8.6\times10^{12}$ &  0.2\\
      B & $1.9\times 10^{5}$ & $1.9\times 10^{-4}$ & $0.8\times 10^{-4}$ &
      $1.8\times10^{49}$  & $3.6\times10^{12}$ &  0.2\\
      C & $1.8\times 10^{6}$& $4.2\times 10^{-4}$ & $0.7\times 10^{-4}$ & 
      $1.8\times10^{49}$ & $4.1\times10^{12}$
      & 0.1\\
      \hline
    \end{tabular}
  \end{center}
\end{table*}

In this subsection we investigate whether circumbinary BHBH accretion
disks could explain simultaneous GW and EM signals of the type
GW150914 and GW150914-GBM. For the BHBH-disk model to explain the
GW150914-GBM event, the following minimum set of requirements must be
met: i) the accretion rate has to be high enough to explain to
observed luminosity, ii) the densities have to be sufficiently low for
``dynamical friction'' not to alter the BHBH inspiral and hence the
waveforms, and iii) the model has to explain why Fermi GBM did not see any
EM signal {\it before} merger.

Point i) is trivial to account for within the BHBH-disk models
described here, utilizing our allowed scale freedom. We scale the BHBH
ADM mass in our simulations to correspond to the event GW150914, i.e.,
$65M_\odot$. Therefore, the individual BH masses become $M_{\rm BH,1}
= 36M_{\odot}$, and $M_{\rm BH, 2} = 29M_{\odot}$. In addition, we
scale the maximum rest-mass density in the disk so that the EM
luminosity in our models matches the inferred equivalent isotropic
luminosity of $\sim 1.8\times 10^{49}$ erg s$^{-1}$ for the
GW150914-GBM event~\cite{connaughton}, assuming a beaming angle of 20
degrees (the approximate jet opening angle in our simulations) and a
$10\%$ conversion efficiency to observable photons (see
Appendix~\ref{sec:appendixA} for how such scaling is performed). To be
precise we assume that the equivalent isotropic luminosity $L_{\rm iso}$
corresponding to the EM flux ($f$) detected by Fermi GBM is related
to the total Poynting luminosity $L_{\rm EM}$ according to
\begin{equation}
\label{eq:fluxL}
 \frac{L_{\rm iso}}{4\pi D_{L}^{2}(z)}=\frac{\epsilon L_{\rm
     EM}/2}{2\pi \eta_{c}D_{L}^{2}(z)}
\end{equation}
where $\epsilon$ is the fraction of the Poynting luminosity that
becomes hard X-ray photons, $\eta_{c}=1-\cos(\theta)$ (with $\theta$
the jet half-opening angle) is a ``collimation'' factor, which equals
0.06 for a $\theta=$20 degrees, and $D_{L}(z)$ is the luminosity
distance. The factor of $1/2$ in the numerator in the right-hand-side
of Eq.~\eqref{eq:fluxL} arises because Fermi GBM would see only the
jet directed towards it, whereas the luminosity output we compute is
from both jets. Equation \eqref{eq:fluxL} yields
\begin{equation}
\label{eq:LLiso}
L_{\rm EM}=\frac{\eta_{c} L_{\rm iso}}{\epsilon}
\end{equation}
We then set $L_{\rm EM}$ such that $L_{\rm iso}=1.8\times 10^{49} \rm
erg\ s^{-1}$, $\eta_c=0.06$, and $\epsilon=0.1$, a fiducial value.

Table~\ref{tab:t3} shows physical quantities for each of the cases
using this scaling. In the predecoupling phase, the accretion rate
reaches $\sim 10^{-3} M_{\odot} s^{-1}$ for all disk models, and after
merger the rates settle approximately to the same value, $\sim
10^{-4} M_{\odot} s^{-1}$. The maximum rest-mass densities in the
disks are $\sim 10^5$g cm$^{-3}$ for cases A and B, and $\sim 10^6$g
cm$^{-3}$ for case C. Note also, that the maximum attainable value of
the Lorentz factor $\Gamma_L$ for steady-state, axisymmetric jets
approximately equals the quantity
$B^2/(8\pi\rho$)~\cite{B2_over_2RHO_yields_target_Lorentz_factor},
which can reach values of $\gtrsim 100$ within the funnel.

In~\cite{fedrow} it is argued that for the scenario where an accreting
BHBH system is formed through stellar core fragmentation, at gas
densities $\rho\gtrsim 10^{6} - 10^{7}$ g cm$^{-3}$, dynamical
friction between the BHs and the gas will change the coalescence
dynamics and the GW signal in a measurable way. However, in the
relativistic simulations presented in~\cite{fedrow} the initial BHBH
was immersed in a nonrotating, constant density cloud, employing
prototype densities that could arise in an evolved progenitor star. In
other words, the binary did not form self-consistently following
fragmentation inside a massive progenitor star. Nevertheless,
Tab.~\ref{tab:t3} shows that one can explain the GW150914-GBM
luminosity with rest-mass densities $\lesssim 10^{6}$ g cm$^{-3}$,
which according to~\cite{fedrow} would not affect significantly the GW
signal. Our study also suggests that models with thicker tori, such as
those encountered following stellar
collapse~\cite{shibata,Sun:2017voo}, may require lower densities and
stand a better chance at explaining both the GW and the EM signals
from GW150914-GBM.

Another consistency check we must perform is that GW150914-GBM
appeared about 0.4 s following the BHBH merger observed by the LIGO
detectors. However, we find that magnetized BHBH-disk systems have jet
output before merger whose power increases as the binary approaches
merger and after merger. Note that the increasing EM power output as
the orbital separation decreases that we discovered
in~\cite{farris2,gold2}, and is also present here, is consistent with
the findings reported recently in~\cite{Kelly:2017xck}, who studied
BHBH mergers in magnetized clouds. The question then is: if the
BHBH-disk model is able to explain the multimessenger signatures of
the GW150914-GBM event, why did Fermi GBM not observe an EM signal
during the inspiral phase? The fact that GW150914-GBM was a marginal
(3$\sigma$) detection allows for the possibility that the signal
before merger was not sufficiently strong to trigger Fermi
GBM. In~\cite{Paschalidis:2016agf} it was pointed out that the time
delay of about $1000$M$\sim 0.3($M$/65M_\odot)$ s between merger and the
boost in the jet luminosity found in earlier BHBH-disk
simulations~\cite{farris2,gold2} could explain this delay in the
GW150914-GBM signal in the sense that it could be weak enough prior to
merger and grow strong enough to be detectable only after merger.

\begin{figure}
\centering
\includegraphics[scale=0.3]{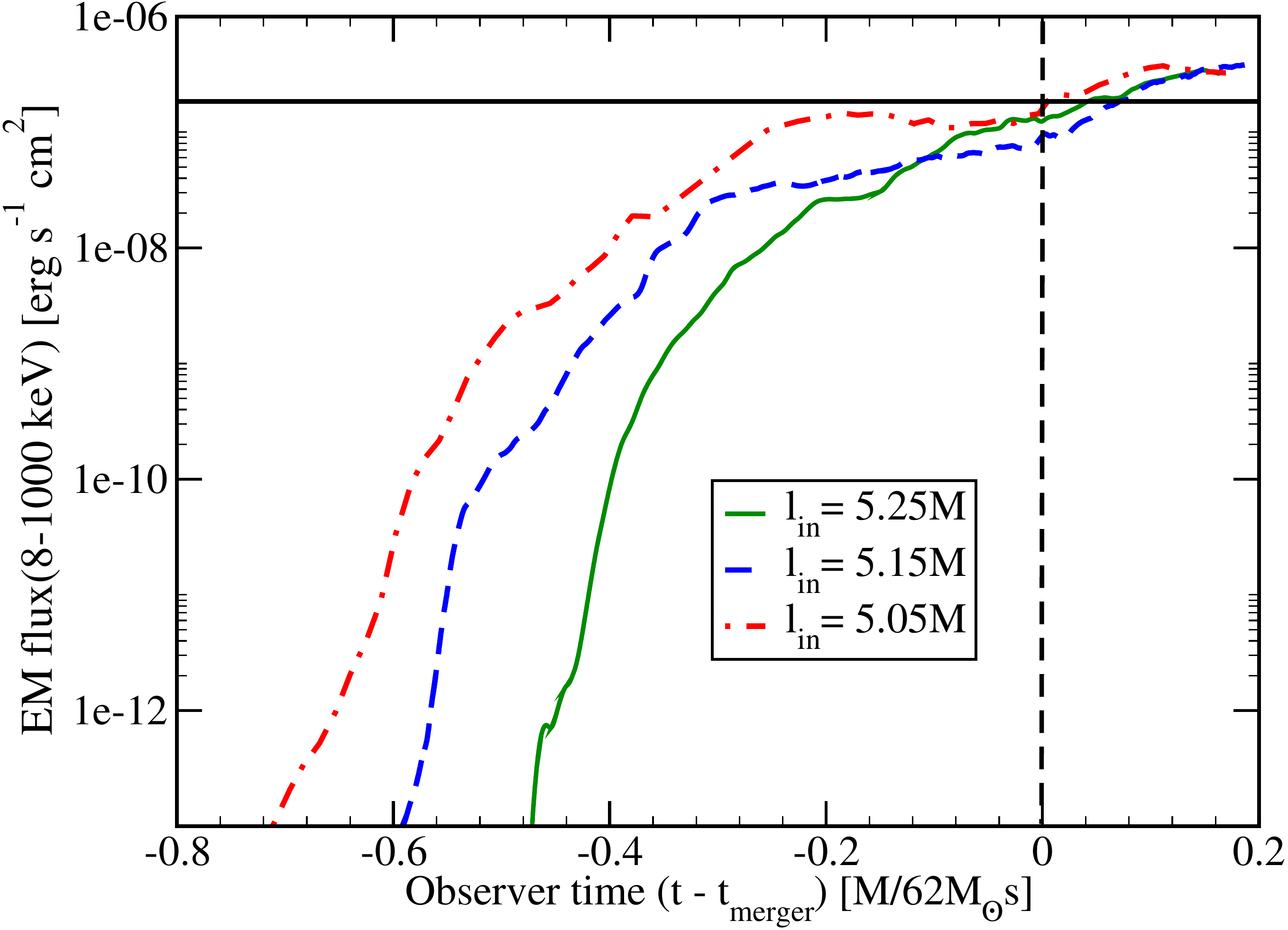}
\caption{GBM flux vs. time for each case assuming the source lies at
  redshift $z=0.09$.  The dashed vertical line is the time at which
  the binary merges as measured by the observer. The horizontal line
  shows the sensitivity of Fermi GBM.
\label{fig:em_detect}}
\end{figure}

To assess this possibility for the models we consider here, we compute
the energy flux Fermi GBM would detect as a function of time as follows
\begin{equation}
\label{eq:flux}
f = \frac{\epsilon L_{\rm EM}/2}{2\pi \eta_{c} D_{L}^{2}(z)}.
\end{equation}
In all estimates, we choose a fiducial value of $\epsilon = 0.1$ and
$\eta_{c}=0.06$, and set the maximum $L_{\rm EM}$ through
Eq.~\eqref{eq:LLiso}.  Figure~\ref{fig:em_detect} shows the EM flux vs
time curve for each case at redshift $z=0.09$. The plot also shows an
estimate for the Fermi GBM 1s sensitivity (horizontal line). Note that
for all cases Fermi GBM would detect the burst occurring post-merger
but not during the inspiral, which is consistent with the
observation~\cite{connaughton}. Notice also that the time delay to
reach the peak luminosity after merger is of order 0.2 seconds, which
is also consistent with GW150914-GBM.  The time delay is due to the
fact that the binary carves out a low density hollow which is roughly
the same for most disk models. Therefore, this time delay, during
which the hollow is filled by accreting gas, is likely universal.

The last aspect one has to account for is the duration of the
burst. In principle, the EM burst duration is likely to be connected
to the engine lifetime which in our scenario is the accretion
disk. The disk accretion time in our models is $O(100\rm s)$, and
hence much longer than the GW150914-GBM burst duration of 1s. This is
because we did not carefully design our disk models to match all
aspects of the scenario considered. The accretion timescale depends on
the net magnetic vertical flux inside the disk which determines the
effective alpha viscosity and the accretion rate (for a fixed
luminosity and fixed efficiency of converting accretion power to EM
power). The timescale also depends on the density distribution in the
disk, the specific angular momentum profile etc. This large parameter
space must be explore in order to be able to find the right disk
model, i.e., the one which yields the desired accretion time. Recent
simulations of collapsing massive stars in full general
relativity~\cite{Sun:2017voo}, find accretion timescales of order 1s
when scaled to $M=65M_\odot$. Moreover, the simulations
of~\cite{Reisswig:2013sqa} that form black hole binaries from
fragmenting, collapsing stars in full general relativity, find
accretion disks with accretion timescales $\sim 0.3$ s, when scaled to
$65M_\odot$. Thus, it is likely that an appropriate disk model can be
found, and a more promissing approach for building the disk model is
to start with matter initial conditions that approximate the tori
found in such self-consistent
calculations~\cite{Sun:2017voo,Reisswig:2013sqa}. The topic requires
further investigation and we plan to tackle it in a future work.

Extrapolating from our recent compact binary simulations modeling
short gamma-ray burst (sGRB) engines~\cite{vasilis,ruizm} and
supermassive star ~\cite{Sun:2017voo}, a unified analytic model was
proposed in~\cite{shapiro} to crudely explain some of the most
significant global parameters characterizing engines formed from
compact binary mergers and stellar collapse. We present a comparison
of our findings with the model of~\cite{shapiro} in
Appendix~\ref{app:B} adopted for GW150914-GBM.

\subsection{Supermassive binary black holes}

In the following, we scale the BHBH ADM mass to equal $M =
10^{8}M_{\odot}$. GWs from such supermassive BHBHs may be detected
ultimately by the spaced based LISA detector~\cite{amaro,2017arXiv170200786A},
targeting GW frequencies $10^{-5}-1 \rm Hz$, as well as Pulsar Timing Arrays
\cite{hobbs,tanaka,sesana}, targeting GW frequencies $10^{-9} -
10^{-6}\rm Hz$.  The GW strain for a $q=29/36$ mass ratio binary at
orbital separation $d=10M$ is $h\sim 10^{-16}(M/10^8M_\odot)(D_L/6.7
\rm Gpc)^{-1}$, where $D_L$ is the luminosity distance and the
normalization chosen corresponds to redshift $z\simeq 1$ in a standard
$\Lambda$CDM cosmology. The corresponding GW frequency is $f_{\rm GW}
\sim 6.5\times 10^{-5} \bra{M/10^8M_\odot}^{-1}[(1+z)]^{-1}$Hz, with
merger frequency $f_{\rm GW, merger} \sim 6.5\times 10^{-4}
\bra{M/10^8M_\odot}^{-1}[(1+z)]^{-1}$Hz. The merger time from $d=10M$
is $\sim 4(M/10^8M_\odot)(1+z)$d. This GW strain is above the LISA
sensitivity at these frequencies \cite{amaro}, hence these systems may
be detectable by LISA. Moreover, with proper modeling, EM precursor
signals can trigger targeted GW searches in the LISA band with a
substantial lead time, should such systems merge frequently enough.
Conversely, LISA can also localize a source on the sky weeks before
merger, and then wide-field telescopes will have a comfortable lead
time to monitor the area~\cite{kocsis,kocsis2,LangHughes2008}.
However, we note that binaries with mass $\sim 10^6-10^7M_\odot$ are more
promising LISA sources. On the other hand binaries with masses $10^9 -
10^{10}M_\odot$ are promising sources for Pulsar Timing
Arrays~\cite{hobbs,tanaka,sesana}.

We now also scale the maximum rest-mass density in the disks to be
$\rho_{0} = 5.6\times 10^{-11}$ g cm$^{-3}$, which gave a Poynting
luminosity near the Eddington limit in~\cite{gold}. Table~\ref{tab:t5}
shows that the accretion rates found here are in the range
$10^{-7}-10^{-9} M_\odot$ s$^{-1}$ but the Poynting luminosity is near
the Eddington luminosity. Only case C we consider deviates by an order
of magnitude from the Eddington luminosity.

\begin{table*}[th]
  \caption{Table of values for the maximum rest-mass density
    $\rho_{0}^{\rm max}$, disk-BH mass ratio
    $M_{\text{disk}}/M_{\text{BH}}$, accretion rate $\dot{M}$,
    luminosity $L$ in units of the Eddington luminosity $1.26\times
    10^{46}(M/10^8M_\odot)$ (erg s$^{-1}$), the magnetic-field
    strength measured by an observer comoving with the plasma averaged
    over a cube of side equal to the remnant coordinate BH apparent
    horizon radius $r_{\rm AH}$ and located immediately above the
    remnant BH at the end of the simulation, and the time
    $\tau_{\text{delay}}$ between merger and peak luminosity. Note
    that the quantities listed in the table scale with the BH mass and
    density as listed in the footnotes.
    \label{tab:t5}}
  \begin{center}
    \begin{tabular}{ccccccc} 
      \hline
      \multicolumn{7}{c}{Simulation Results (Supermassive BHBH Scaling) }\\
      \hline
      \hline
      Case & $\rho_{0}^{\text{max}}$ g cm$^{-3}$ & $M_{\text{disk}}/M_{\text{BH}}$\footnote{$M_{\text{disk}}/M_{\text{BH}} \propto
    (\rho_0^{\rm max}/5.6\times 10^{-11}\rm g\ cm^{-3})(M_{\rm
      BH}/10^8M_\odot)^2$} & 
      $\dot{M}$ ($M_{\odot}$ s$^{-1}$)\footnote{$\dot M \propto (\rho_0^{\rm max}/5.6\times
    10^{-11}\rm g\ cm^{-3})(M_{\rm BH}/10^8M_\odot)^2$} & $L_{\rm EM}/L_{\rm edd}$\footnote{$L_{\rm
      EM}/L_{\rm edd} \propto (\rho_0^{\rm max}/5.6\times 10^{-11}\rm
    g\ cm^{-3})$} & 
      $B_{p}$ (Gauss)\footnote{$B_p \propto (\rho_0^{\rm max}/5.6\times 10^{-11}\rm
    g\ cm^{-3})^{1/2}$} &  $\tau_{\rm delay}/(1+z)$\footnote{$\tau_{\rm delay}/(1+z) \propto
    (M/10^8M_\odot)$}\\
      \hline
      A & $5.6\times 10^{-11}$& $6.7\times 10^{-7}$ & $2.0\times 10^{-7}$ & 
      $0.9$ & $1.3\times10^{5}$ & 3.6 d
      \\
      B & $5.6\times 10^{-11}$ & $1.4\times 10^{-7}$ & $6.2\times 10^{-8}$ &
      $1.0$  & $6.1\times10^{4}$ & 3.6 d \\
      C & $5.6\times 10^{-11}$& $3.4\times 10^{-8}$ & $6.0\times 10^{-9}$ & 
      $0.1$ & $2.3\times10^{4}$ & 1.8 d\\
      \hline
    \end{tabular}
  \end{center}
\end{table*}

For all models, the characteristic effective temperature in the bulk
of the disk is approximately insensitive to the disk model or inspiral
epoch and is similar to what we reported in~\cite{gold}
and~\cite{gold2}:
\begin{equation}
  T_{\rm eff} \sim 10^{5}\left(\frac{L_{\rm EM}}{L_{\rm Edd}}\right)^{1/4}
  \left(\frac{M}{10^{8}M_{\odot}}
\right)^{-1/4}{\rm K},
\end{equation}
where we assumed $\rho_{0}\epsilon = aT^{4}$, to estimate the local
temperature, and we set $\rho_0$ as in~\cite{gold}. Note that these
temperatures are only crude estimates, because we do not account for
microphysics here. In our calculations, the temperature is also
position dependent and is close to $6\times10^{5}
(M/10^{8}M_{\odot})^{-1/4}$K near the inner disk edge and goes to $2-3 \times
10^{5} (M/10^{8}M_{\odot})^{-1/4}$K in the bulk of the disk. So, the order of
magnitude is $10^5$K. In some recent Newtonian calculations with more
detailed radiation microphysics (e.g.~\cite{Farris2015MNRAS.447L..80F,
  Tang:2018rfm}) the temperatures obtained are somewhat higher. But
those calculations are all in 2D and based on alpha-disk viscosity
laws and not on 3D GRMHD as performed here, with magnetic fields
providing the effective turbulent viscosity, and this may account for
the difference. The corresponding characteristic thermal radiation
frequencies ($\nu_{\rm bb} \sim k_B T_{\rm eff}/h$) are given by
\begin{equation}
  \nu_{\rm bb} \sim  10^{15} \bra{\frac{M}{10^8M_\odot}}^{-1/4} \bra{\frac{L_{\rm b}}{L_{\rm Edd}}}^{1/4}(1+z)^{-1} \rm Hz.
\end{equation}
Current and future EM detectors such as PanStarrs \cite{kaiser}, the
LSST \cite{abell} and WFIRST \cite{green} could detect such thermal EM
signals.


\section{Conclusions}\label{sec:conclusions}

We performed MHD simulations of binary black holes with mass ratio
$q=29/36$ that accrete magnetized matter from a circumbinary accretion
disk. We consider three initial disk models that differ in their scale
heights, physical extent and in their magnetic field content in order
to test whether previous properties of MHD accretion flows onto binary
black holes are sensitive to the initial disk model. We find that the
presence of periodicities in the accretion rate, the emergence of
jets, the time delay between merger and the boost in the jet
luminosity that we previously discovered~\cite{farris2,gold,gold2} are
all robust features and largely insensitive to the choice of initial
disk model.

As in our previous studies we ignored the disk self-gravity and
adopted a simplified $\Gamma$-law EOS ($\Gamma=4/3$), which allows us
to scale the binary black hole mass and disk densities to arbitrary
values. Thus, our results have implications both for LIGO black-hole
binaries and for supermassive black hole binaries at centers of
luminous AGNs and quasars.

Scaling our simulations to LIGO GW150914 we find that magnetized disk
accretion onto binary black holes could explain both the GWs detected
from this system, and the EM counterpart GW150915-GBM reported by the
Fermi GBM team 0.4s after LIGO's GW150915. When scaling to
supermassive black hole binaries, we find that at late times flow
properties, temperatures, thermal frequencies, are all
robust displaying only modest dependence on the disk model.
Nevertheless, the range of disk thickness and ratio of magnetic-to-gas pressure in
our survey is limited by what we can achieve with current
computational resources and methods. As computational resources grow
and numerical techniques advance we will be able to probe wider ranges
of these parameters.

Using the remnant BH-disk parameters found for the simulations, we
also tested the predictions of the unified model presented
in~\cite{shapiro} for gamma-ray bursts formed from compact binary
mergers and massive star collapse. We found that the disk accretion
rate, Poynting luminosity, EM radiation efficiency and the disk
lifetimes are in rough agreement between the model and the
simulations.

\appendix
\section{Scaling with Rest-Mass Density}
\label{sec:appendixA}

Our model may be scaled to arbitrary rest-mass density $\rho_0$
because we neglect the self-gravity of the disk and we adopt a
$\Gamma$-law EOS. The density everywhere is determined by specifying
its maximum value, $\rho_{0}^{\rm max}$, which can be set by requiring
that once the remnant black hole and disk achieve quasistationary
equilibrium, $L_{\rm EM} = L_{\rm EM\text{ (observed)}}$.
Quantities such as the accretion rate and EM luminosity are provided
in code units, and then $\bar{\rho}_{{\rm max}}$ is rescaled to
establish the luminosity condition and $M_{\rm BH}$ to equal the
desired binary ADM mass. Here and below we consider barred quantities
to be given in code units.  Knowing how all other quantities scale
with $\rho_{0}^{\rm max}$ and $M_{\rm BH}$ then allows us to rescale
their values appropriately. Below we show how their scaling with
$\rho_{0}^{{\rm max}}$ and $M_{\rm BH}$ is established.

The accretion rate $\dot{M}$ onto the merger remnant black hole scales
as
\begin{equation}
\dot{M} \sim \rho_{0} M_{BH}^{2} \Rightarrow \frac{\dot{M}_{\rm BH}}{\bar{\dot{M}}_{\rm BH}}
                =\frac{\rho_{0}}{\bar{\rho}_{0}} \left( \frac{M_{BH}}{\bar{M}_{BH}} \right)^2.      
\label{eq:mdotscale}
\end{equation}
To show that $P_{\rm gas}$ must scale with $\rho_0$, we begin
with the fact that we adopt a $\Gamma$-law EOS,
\begin{align*}
P_{\rm gas} &= (\Gamma - 1)\rho_{0}\epsilon \\
  &= (\Gamma - 1)\rho_{0}\left(h - 1 - \frac{P}{\rho_{0}}\right)
\end{align*}
where $\epsilon$ is the specific internal energy, and $h = 1 + \epsilon + P/\rho_{0}$ is the 
specific enthalpy. Now using $P_{\rm gas}=K\rho_{0}^{\Gamma}$, we solve for $P_{\rm gas}/\rho_{0}$ to 
get
\begin{equation}
\frac{P_{\rm gas}}{\rho_{0}} = \frac{\Gamma - 1}{\Gamma}(h - 1)
\label{eq:poverrho}
\end{equation}
As shown in ~\cite{farris3}, $h = h(r, \theta)$ is fixed,
dimensionless, and completely determined by the initial parameters
$R_{\rm in}$, $l_{\rm in}$, and $q$ (see Eq. A12 in
\cite{farris3}). The right-hand-side of Eq.~\ref{eq:poverrho} is then a constant,
and hence
\begin{equation}
\frac{P_{\rm gas}}{\bar{P}_{\rm gas}} = \frac{\rho_{0}}{\bar{\rho}_{0}}
\label{eq:Pscale}
\end{equation}
Now in our simulations, we set the maximum value of the ratio
$P_{\rm mag}/P_{\rm gas}$ at $t=0$, where $P_{\rm mag} = B^{2}/8\pi$.
Hence, to preserve this ratio upon rescaling, we have
\begin{equation}
\frac{B}{\bar{B}} = \left( \frac{\rho_{0}}{\bar{\rho}_{0}}\right)^{1/2}.
\label{eq:bscale}
\end{equation}
Finally, since $L_{\rm EM} \sim B^2 M_{\rm BH}^{2}$, we have
\begin{equation}
\label{Lpoyn}
\frac{L_{\rm EM}}{\bar{L}_{\rm EM}} = \frac{\rho_{0}}{\bar{\rho}_{0}}
					\left( \frac{M_{BH}}{\bar{M}_{BH}} \right)^2.
\end{equation}

Thus when we may set $L_{\rm EM} = L_{EM\text{ (observed)}}$, and use
Eq.~\ref{Lpoyn} to rescale the maximum rest-mass density after
determining $\rho_{0}/\bar\rho_{0}$.  Then we can use
Eqs.~\ref{eq:mdotscale}, \ref{eq:Pscale}, and \ref{eq:bscale} to
compute the remaining quantities.

\section{Comparison with simple estimates}
\label{app:B}

\begin{table*}[th]
	\caption{Order of Magnitude Comparison of Simulation Results with Model in~\cite{shapiro}.\label{tab:stu}}
  \begin{center}
	\begin{tabular}{ccccccccc} 
	  \hline         
	  \hline
        \multirow{2}{*}{Case}  &
      \multicolumn{2}{c}{\underline{\hspace{0.25cm} $L_{\rm EM}/\dot{M}_{\rm eq}$\hspace{0.25cm}} } &
      \multicolumn{2}{c}{\underline{\hspace{0.25cm} $t_{\text{disk}}/M_{\rm BH}$\hspace{0.25cm}} } &
      \multicolumn{2}{c}{\underline{\hspace{0.25cm} $\rho M_{\rm BH}^2$\hspace{0.25cm}} } &
      \multicolumn{2}{c}{\underline{\hspace{0.25cm} $B_p^2 M_{\rm BH}^2$\hspace{0.25cm}} }  \\
	 & \cite{shapiro}    &  Simulations & \cite{shapiro} & Simulations & \cite{shapiro} & Simulations & \cite{shapiro}  & Simulations \\
	\hline
	A & $0.01$ & 0.01  & $10^6$  & $10^6$ & $10^{-14}$ & $ 10^{-12}$ & $ 10^{-12}$ & $ 10^{-13}$ \\
	B & $0.01$ & 0.1  & $10^{5}$  & $10^5$ & $ 10^{-13}$ & $10^{-12}$ & $ 10^{-12}$ & $ 10^{-13}$ \\
	C & $0.01$ & 0.1  & $10^{6}$  & $10^6$ & $ 10^{-13}$ & $10^{-12}$ & $ 10^{-12}$ & $ 10^{-14}$ \\
	\hline
	\end{tabular}
  \end{center}
\end{table*}

Recently a unified analytic model was proposed in~\cite{shapiro} to explain
several key global parameters characterizing an accretion disk--black hole
system that can launch a jet consistent with the BZ
mechanism~\cite{bz}. The model accounted for disk-black hole systems
formed either through compact binary mergers or massive star
collapse. Here we apply the model to the case of GW150914. The
inferred characteristic disk parameters for a viable BH-disk system
for GW150914, which has a low isotropic luminosity ($\sim 10^{49}$ erg
s$^{-1}$), are given in Eq. 19 of~\cite{shapiro} as
\begin{equation}
  \label{GW150914}
  \begin{split}
    &\frac{M_{\rm disk}}{M} \sim 1\times 10^{-5}, 
    \ \ \frac{R_{\rm disk}}{M} \sim 20, \\
    &{\dot M_{\rm eq}} \sim 0.9 \times 10^{-3}~M_{\odot}~{\rm s^{-1}}.
  \end{split}
\end{equation}

Here $R_{\rm disk}$ is roughly identified with the outer radius of the
disk.  The inferred values are determined by matching the model to the
observed luminosity {\it and} lifetime of GW150914-GBM, ignoring
beaming and off-axis viewing.

The value for the accretion rate is within an order of
magnitude of the value our simulations predict here in order to match
the luminosity of GW150914-GBM (see Tab.~\ref{tab:t3}).

In geometrized units the model of~\cite{shapiro} generally predicts
\begin{equation}\dot{M}_{\text{eq}}\sim 4
\left(\frac{M_{\text{disk}}}{M}\right)
\left(\frac{M}{R_{\text{disk}}}\right)^{3},\end{equation}
and
\begin{equation}L_{BZ}\sim \frac{1}{10}
\left(\frac{M_{\text{disk}}}{M}\right)
\left(\frac{M}{R_{\text{disk}}}\right)^{3}
\left(\frac{a}{M}\right)^{2},\end{equation}
and hence
\begin{equation}\frac{L_{BZ}}{\dot{M}_{\rm eq}} \sim \frac{1}{40}
\left(\frac{a}{M}\right)^{2},\end{equation}
(Eqns. 11-13 of~\cite{shapiro}).

The model also predicts a characteristic disk density and the magnetic
field strength near the BH poles:
\begin{equation}
\rho M^2 \sim \frac{1}{\pi}\left(\frac{M_{\rm disk}}{M}\right)\left(\frac{M}{R_{\rm disk}}\right)^3,
\end{equation}
\begin{equation}
B_p^2 M^2 \sim 8\left(\frac{M_{\rm disk}}{M}\right)\left(\frac{M}{R_{\rm disk}}\right)^3
\end{equation}
(Eqs. 9 and 10 of~\cite{shapiro}).

The model also predicts for the disk lifetime
\begin{equation}t_{\text{disk}}/M \sim \frac{(M_{\rm disk}/M)}{\dot{M}_{\rm eq}} 
\sim \frac{1}{4}
\left(\frac{R_{\text{disk}}}{M}\right)^{3}
\end{equation}

In table~\ref{tab:stu} we compare results from our simulations to
those predicted by the model in~\cite{shapiro}. Within one to two
orders of magnitude, these predictions are consistent with the results
of our simulations. This is within the accuracy of the model and its
simplified treatment.


\acknowledgements We thank the Illinois Relativity Group REU team
members Eric Connelly, Cunwei Fan, John Simone and Patchara
Wongsutthikoson for assistance in creating Fig.~\ref{fig:visit}.  We
also thank Roman Gold for sharing his scripts that helped in
generating Fig. 2. This work has been supported in part by NSF Grants
PHY-1602536 and PHY-1662211, and NASA Grants NNX13AH44G and
80NSSC17K0070 at the University of Illinois at Urbana-Champaign. VP
gratefully acknowledges support from NSF grant PHY-1607449, NASA grant
NNX16AR67G (Fermi) and the Simons foundation. This work used the
Extreme Science and Engineering Discovery Environment (XSEDE), which
is supported by NSF grant number OCI-1053575. This research is part of
the Blue Waters sustained-petascale computing project, which is
supported by the National Science Foundation (award number OCI
07-25070) and the state of Illinois. Blue Waters is a joint effort of
the University of Illinois at Urbana-Champaign and its National Center
for Supercomputing Applications.

\bibliographystyle{apsrev4-1}        
\bibliography{bhbh_disk}

\end{document}